\documentclass[a4paper,twocolumn,prb,showpacs,superscriptaddress,amsmath,amssymb,aps]{revtex4-2}
\usepackage[T1]{fontenc}
\usepackage{verbatim}
\usepackage{amsmath}
\usepackage{amssymb}
\usepackage{graphicx}
\usepackage[colorlinks,linkcolor=blue,anchorcolor=blue,citecolor=blue,urlcolor=blue]{hyperref}
\usepackage{siunitx}

\graphicspath{{figure/}{figsgaoerb/}} % 读取图片从figs/文件夹中读取，而不是当前文件夹

\makeatletter

\usepackage{epstopdf}
\usepackage{ulem}
\usepackage{soul}
\usepackage{color}
\newcommand{\ignore}[1]{}

\makeatother

\begin{document}
	%\title{The Role of Disorder in Two-Dimensional Dirac Fermions}
	\title{Disorder Effects on the Quasiparticle and Transport Properties of Two-Dimensional Dirac Fermionic Systems}
	\author{Bo Fu}
	\thanks{These authors contributed equally to this work}
	\affiliation{International Center for Quantum Design of Functional Materials (ICQD), Hefei National Research Center for Physical Sciences at the Microscale, University of Science and Technology of China, Hefei, Anhui 230026, China}
\affiliation{School of Sciences, Great Bay University, Dongguan, China}
	\author{Yanru Chen}
	\thanks{These authors contributed equally to this work}
	\affiliation{International Center for Quantum Design of Functional Materials (ICQD), Hefei National Research Center for Physical Sciences at the Microscale, University of Science and Technology of China, Hefei, Anhui 230026, China}
	\affiliation{Hefei National Laboratory, University of Science and Technology of China, Hefei, Anhui 230088, China}
	\author{Weiwei Chen}
	\affiliation{Institute of Natural Sciences, Westlake Institute for Advanced Study, 18 Shilongshan Road, Hangzhou 310024, Zhejiang Province, China}
	\author{Wei Zhu}
	\affiliation{Institute of Natural Sciences, Westlake Institute for Advanced Study, 18 Shilongshan Road, Hangzhou 310024, Zhejiang Province, China}
	\author{Ping Cui}
	\affiliation{International Center for Quantum Design of Functional Materials (ICQD), Hefei National Research Center for Physical Sciences at the Microscale, University of Science and Technology of China, Hefei, Anhui 230026, China}
	\affiliation{Hefei National Laboratory, University of Science and Technology of China, Hefei, Anhui 230088, China}
	\author{Qunxiang Li}
	\thanks{Corresponding author. E-mail: liqun@ustc.edu.cn}
	\affiliation{International Center for Quantum Design of Functional Materials (ICQD), Hefei National Research Center for Physical Sciences at the Microscale, University of Science and Technology of China, Hefei, Anhui 230026, China}
	\affiliation{Hefei National Laboratory, University of Science and Technology of China, Hefei, Anhui 230088, China}
	\author{Zhenyu Zhang}
	\thanks{Corresponding author. E-mail: zhangzy@ustc.edu.cn}
	\affiliation{International Center for Quantum Design of Functional Materials (ICQD), Hefei National Research Center for Physical Sciences at the Microscale, University of Science and Technology of China, Hefei, Anhui 230026, China}
	\affiliation{Hefei National Laboratory, University of Science and Technology of China, Hefei, Anhui 230088, China}
	\author{Qinwei Shi}
	%\thanks{Corresponding author. E-mail: phsqw@ustc.edu.cn}
	\affiliation{International Center for Quantum Design of Functional Materials (ICQD), Hefei National Research Center for Physical Sciences at the Microscale, University of Science and Technology of China, Hefei, Anhui 230026, China}
	\date{\today}
	\begin{abstract}
		Despite extensive existing studies, a complete understanding of the role of disorder in affecting the physical properties of two-dimensional Dirac fermionic systems remains a standing challenge, largely due to obstacles encountered in treating multiple scattering events for such inherently strong scattering systems. Using graphene as an example and a nonperturbative numerical technique, here we reveal that the low energy quasiparticle properties are considerably modified by multiple scattering processes even in the presence of weak scalar potentials. We extract unified power-law energy dependences of the self-energy with fractional exponents from the weak scattering limit to the strong scattering limit from our numerical analysis, leading to sharp reductions of the quasiparticle residues near the Dirac point, eventually vanishing at the Dirac point. The central findings stay valid when the Anderson-type impurities are replaced by correlated Gaussian- or Yukawa-type disorder with varying correlation lengths. The improved understanding gained here also enables us to provide better interpretations of the experimental observations surrounding the temperature and carrier density dependences of the conductivity in ultra-high mobility graphene samples. The approach demonstrated here is expected to find broad applicability in understanding the role of various other types of impurities in two-dimensional Dirac systems.
	\end{abstract}
	
	\maketitle
	
	\setulcolor{red} 
	\setstcolor{blue} 
	\sethlcolor{yellow} 
	
	\section{Introduction}
	The role of disorder in two-dimensional Dirac fermionic systems \cite{Fisher} was intensively explored in the
	early 1990's, in part motivated by the observations of localized  states in $ d $-wave superconductivity of cuprate
	superconductors \cite{Patrick} and plateau transitions in integer quantum Hall effect \cite{Ludwig}.
	In those pioneering studies, it has been shown that even
	impurities with weak scattering strengths have non-perturbative effects on the quasiparticle properties near the Dirac point \cite{Ludwig,Nersesyan,WenXiaogang},
	resulting in intriguing new physical consequences. As an example, an exact conformal field theory
	was developed to successfully describe the contributions from multiple impurity scattering processes if the nature
	of weak disorder preserves the continuous chiral symmetry \cite{Ludwig,Nersesyan,WenXiaogang,Sbierski2020prx}.
	In this scenario, the electron density of states was shown to possess a power-law
	dependence on energy with fractional exponents, instead of logarithmic behaviors obtained
	within perturbative treatments \cite{Fradkin,Nersesyan2}. Those
	findings not only enriched our physical understanding about such disordered systems,
	but also highlighted the importance of multiple
	impurity scattering processes. Nevertheless, the conformal field theory is not applicable to
	scalar type of impurities that breaks continuous chiral symmetry. 
	In such cases, the elastic scattering time $\tau$ is short, and the Dirac fermionic systems easily enter into the strong scattering limit (the dimensionless parameter $E_{f}\tau/\hbar\leq1$) as the Fermi energy $E_{f}$ approaches the Dirac point
	(here $E_{f}$ is measured relative to
	the Dirac point), calling for new theoretical treatments. 
	
	Separately, since the experimental discovery
	of graphene \cite{Novoselov}, the past decade has seen a substantial rejuvenation of interest in the study of the
	role of disorder in two-dimensional Dirac fermionic systems. In particular, graphene serves as an ideal platform for studying disorder effects at or close to the Dirac point, because the system displays linear dispersion over a large energy range. Indeed, extensive unusual 
	transport properties have been reported using ultrahigh-mobility samples
	\cite{Bolotin,Du,Ponomarenko,Morozov,Dean,Zomer,Mayorov2012,Nuno,kazi2018,WangLujun}, including that the minimum conductivity at the Dirac point strongly depends on temperature, the conductivity is sublinear in carrier density very close to the Dirac point, and there exists a critical carrier density separating a nonmetallic and a metallic regime characterized by the temperature dependence of resistivity.
	Such novel transport behaviors not only reflect intriguing physics around the Dirac point in weakly disordered graphene, but potentially also highlight the importance of exotic disorder effects.  To date, those unconventional transport properties of graphene remain to be fully understood, in part because prevailing theoretical treatments have various limitations. For example, the standard Boltzmann transport theory treatments \cite{Neto,DasSarma,DasSarma2013} only capture subsets of scattering events. % in term of the relaxation time computed by the Fermi-golden rule , although the perturbation calculation \cite{Aleiner} points out again the limitation of perturbation approaches such as the self-consistent Born approximation (SCBA) \cite{ZhengSCBA,Ostrovsky}.
	Several analytical approaches have also been developed to study the role
	of disorder in graphene, as examplified by the functional renormalization-group
	(fRG) approach \cite{Katanin,Sbierski}, but these new developments again only considered the contributions
	of some subsets of multiple impurity scattering processes and still failed to properly describe
	the quasiparticle behavior around the Dirac point. Therefore, it is still a standing
	challenge to reliably treat multi-scattering events in the presence of physically realistic
	disorder without continuous chiral symmetry. An enabling theoretical approach is needed
	to include all the multiple scattering events in order to capture the underlying disorder physics,
	especially near the Dirac point.

	In this work, we present numerically exact results of the quasiparticle and transport properties
	of disordered two-dimensional Dirac fermionic systems as obtained using an accurate momentum-space
	Lanczos method  \cite{zhu2010,Fu}, with disordered graphene around the Dirac point as a concrete example. 
	As shown recently, this method is able to rigorously treat all multiple scattering
	events from random scalar disorder potentials or other types of impurities.
	Strikingly, we extract from numerical data a universal power-law functional
	form of the self-energy in describing the multiple scattering effect of disorder
	on the quasiparticle behavior, which is valid from the weak to strong
	scattering limit. The newly established universal power law enables us to
	further reveal the novel quasiparticle behaviors near the Dirac point, such as the
	unusual energy dependence of the quasiparticle residue.
	We are also able to reproduce the experimentally
	observed conductivity versus the carrier density at different temperatures \cite{Bolotin,Ponomarenko,Dean,Mayorov2012},
	thereby attesting that a proper account of multiple impurity scattering processes is essential
	in understanding the transport properties of disordered graphene.
	The approach demonstrated here is expected to find broad applicability in
	other disordered systems where multiple impurity scattering events play a decisive role.
	
	This paper is organized as follows. The model and methodologies are introduced in Sec. \ref{sec.model}, followed by the numerical results for the self-energy and quasiparticle properties in Sec. \ref{sec.self}. 
	The transport properties are given in Sec. \ref{sec.transport}. 
	We discuss two kinds of correlated impurity: Gaussian- and Yukawa-type disorder in Sec.~\ref{sec.corrlated potential}. 
	Finally, in Sec. \ref{sec.discussion}, we draw some conclusions from our main results.
	
	\begin{figure}[t!]
		\includegraphics[width=8.5cm]{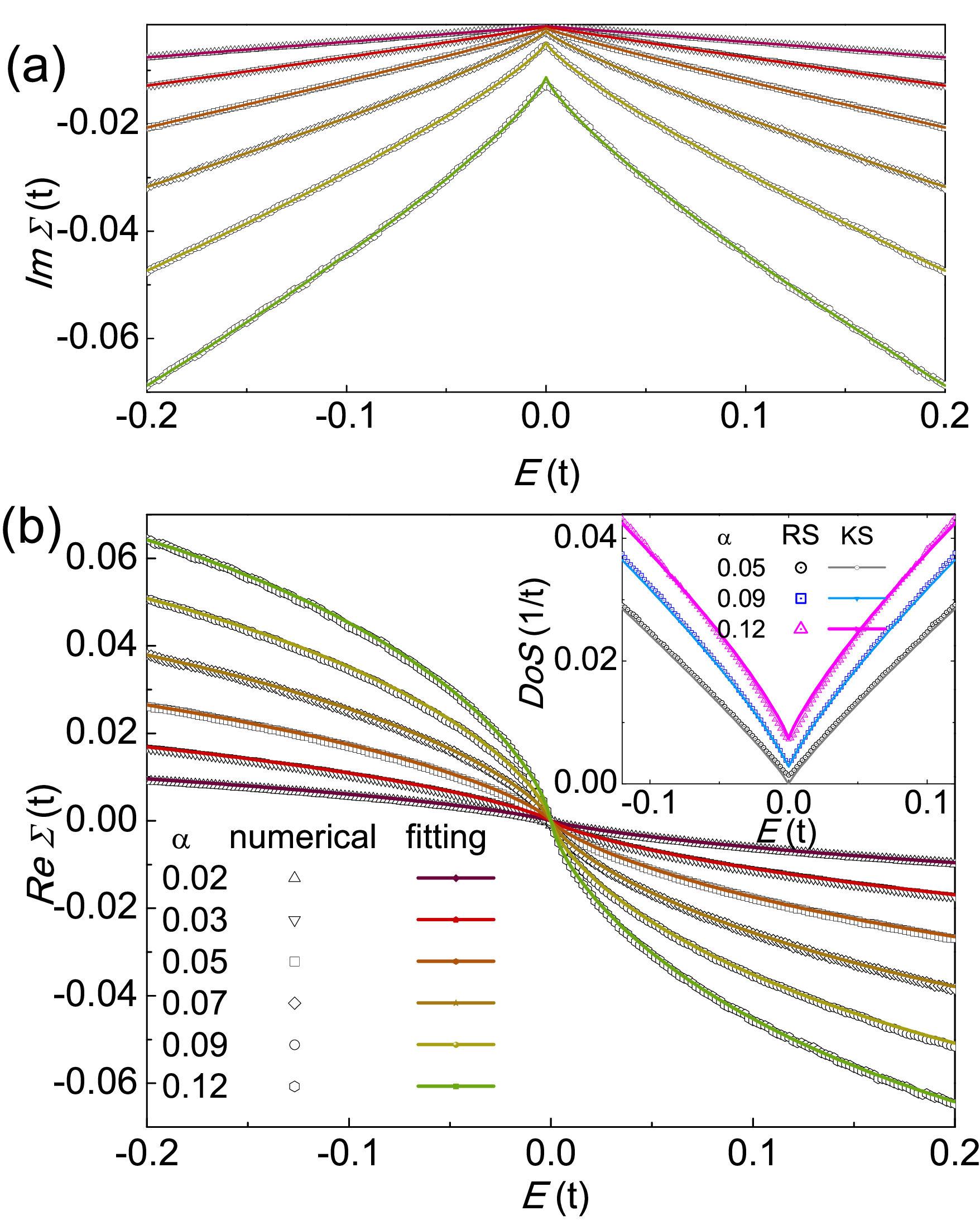} \caption{(a) Imaginary and (b) real parts of the self-energy for disordered graphene with different disorder strengths $(0.02\leqslant\alpha\leqslant0.12)$. 
			The open symbols are the numerical results, while the solid lines are the fitting curves by Eqs.~(\ref{Eq:imaginaryselfenergy}) and ~(\ref{Eq:sigma1}). {The inset in (b) shows the comparison of the density of states (DoS) obtained by the real-space Lanczos method (RS, open symbols) with the calculated results based on the fitted self-energy in momentum space (KS, solid lines) for $\alpha=0.05$, $0.09$, and $0.12$.}}
		\label{fig:self}
	\end{figure}
	
	\section{Model and method}
	\label{sec.model}
	In the absence of disorder, graphene can be modeled by a $\pi$-band tight-binding Hamiltonian. In our calculations,
	the short range Anderson-type disorder is introduced by the on-site energy distributed uniformly and independently within $[-W/2,W/2]$. So we consider the following Hamiltonian on a honeycomb lattice:
	\begin{align}
		H=t\sum_{<ij>}|i\rangle\langle j|+\sum_i V_i |i\rangle\langle i|,
	\end{align}
 	where \textit{t} is the hopping energy between the nearest neighbouring carbon atoms. A dimensionless parameter $\alpha=\frac{A_{c}W^{2}}{12(\hbar v_{f})^{2}\pi}$
	is defined to characterize the strength of uncorrelated Anderson disorder, where
	$A_{c}=\frac{3\sqrt{3}}{2}a^{2}$ is the area of the unit cell, $a$ is the C-C distance,
	and $v_{f}=3at/2\hbar$ the bare group velocity for clean graphene. To explore the quasiparticle properties of disordered graphene we choose a large graphene sample containing millions of atoms $(L^{2}=10000^{2})$ to calculate its retarded self-energy ($\Sigma$) by the momentum-space Lanczos recursive method \cite{zhu2010,Fu}. The large sample in our calculations allows us to choose a small artificial cutoff $\eta=0.001$ to simulate the infinitesimal imaginary energy so that we can extract the self-energy function with high-energy resolution.
	
	\section{Self-energy and quasiparticle properties}
	\label{sec.self}

	\begin{figure}[t!]
		\includegraphics[width=8.5cm]{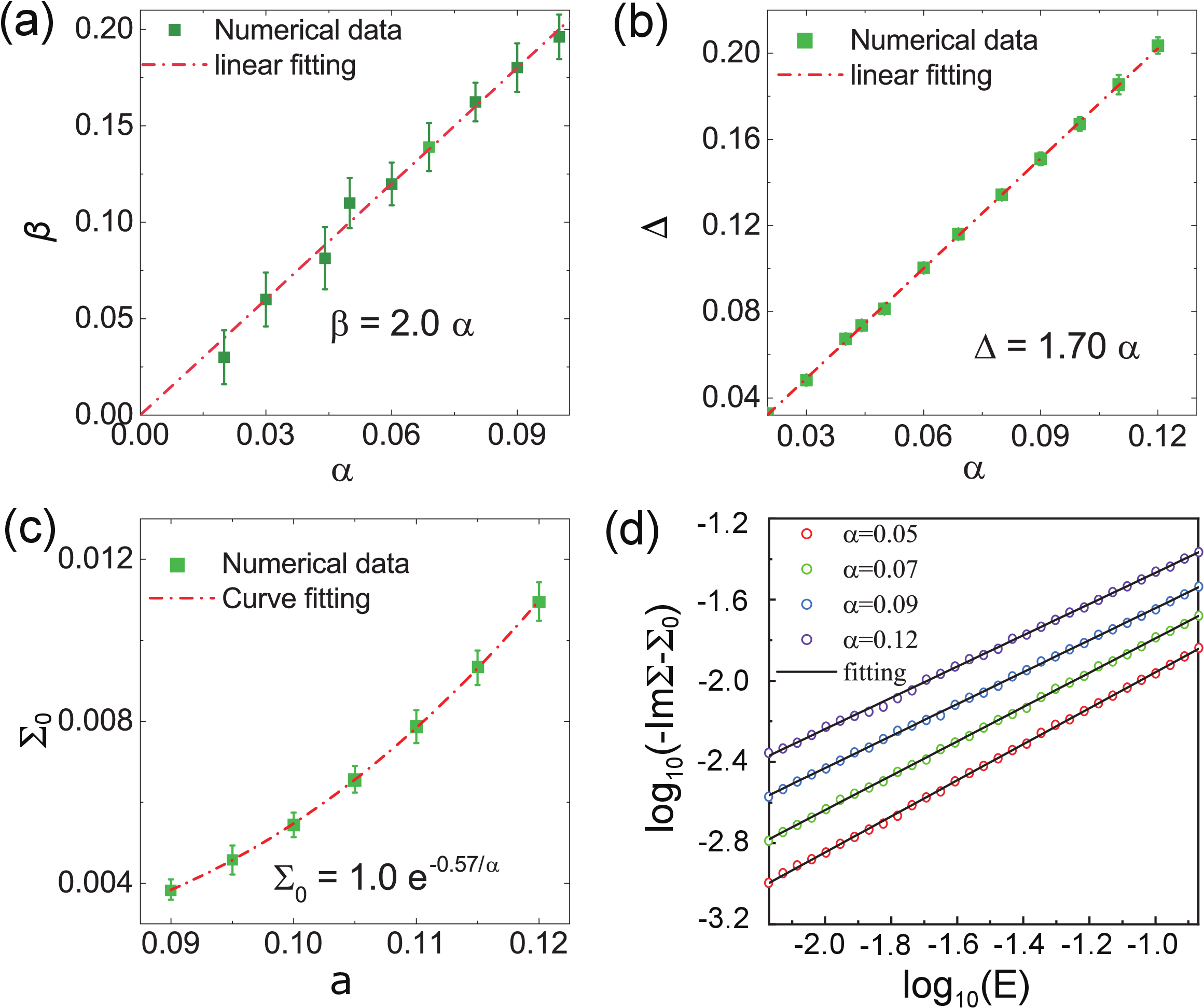}\caption{Numerical fittings of the parameters (a) $\beta$, (b) $\Delta$ and (c) $\Sigma_{0}$ as functions of the disorder strength $\alpha$. $\beta$, $\Delta$ and $\Sigma_{0}$ are obtained from the power-law self-energy fitting in Fig.~\ref{fig:self}. (d) Log-log plot of the imaginary part of the self-energy and the power-law fittings by subtracting the values at the zero energy for several disorder strengths.}
		\label{fig:fitting}
	\end{figure}

	The imaginary part of the self-energy ($\mathrm{Im}\Sigma(E)$) for disordered graphene
	with different disorder strengths ($\alpha$) is shown in Fig.~\ref{fig:self}(a). One can see that, as the disorder strength increases, $\mathrm{Im}\Sigma(E)$ gradually deviates from a linear behavior, and the absolute value of $\mathrm{Im} \Sigma(0)$ increases accordingly. These characteristic features inspire us to use a power-law formula to fit our numerical results, given as
	\begin{equation}
		\begin{aligned}\mathrm{Im}\Sigma(E)=-\Sigma_{0}-\Delta|E|^{1-\beta}\quad\quad0<\beta<1,\end{aligned}
		\label{Eq:imaginaryselfenergy}
	\end{equation}
	
	\noindent where $\beta$, $\Sigma_{0}$ and $\Delta$ are the fitting parameters only determined by the disorder strength $\alpha$. As shown in Fig.~\ref{fig:self}(a), the agreement between the self-energy function form in Eq.~(\ref{Eq:imaginaryselfenergy}) and numerical results is excellent within the low-energy window of $[-0.2t,0.2t]$. Eq.~(\ref{Eq:imaginaryselfenergy}) is further confirmed by the log-log plot of the imaginary part of the self-energy as a function of energy as shown in Fig.~\ref{fig:fitting}(d). More remarkably, only via the Kramers-Kronig relation, we identify the functional
	form of the real part of the self-energy ($\mathrm{Re} \Sigma(E)$) without introducing any other adjustable parameter except the high energy cutoff as
	\begin{equation}
		\begin{aligned}\mathrm{Re}\Sigma(E)=D\mathrm{sgn}(E)|E|^{1-\beta}+CE,\end{aligned}
		\label{Eq:sigma1}
	\end{equation}
	
	\noindent where sgn($E$) is the signum function, $C=\frac{2E_{c}^{-\beta}}{\pi\beta}\Delta$,
	and $D=-\mathrm{cot}(\frac{\pi}{2}\beta)\Delta$. The high energy cutoff is chosen as $E_{c}\approx2.7t$, which has the same order of magnitude of the band width.
	Such a functional form in Eq.~(\ref{Eq:sigma1}) can well fit the numerical results of $\mathrm{Re}\Sigma(E)$,
	as shown in Fig.~\ref{fig:self}(b). This further confirms the correctness of our proposed power-law
	formula for the imaginary part of the self-energy. We have also calculated the spectral function as shown in Appendix.~\ref{subsec.SF}, demonstrating that this power-law relation significantly renormalizes the quasiparticle properties around the Dirac point. Moreover, the inset in Fig.~\ref{fig:self}(b)
	plots the comparison of the density of states (DoS) obtained by the widely used Lanczos method in real space \cite{zhu2009} with that calculated using our fitted self-energy, showing again the perfect agreement with each other. More discussions about the DoS are given in Appendix.~\ref{subsec.DoS}.
	
	Equations (\ref{Eq:imaginaryselfenergy}) and (\ref{Eq:sigma1}) are the main discoveries of the work \cite{note1}, reflecting that proper treatments of all orders of multi-scattering events will uncover the novel quasiparticle properties around the Dirac point. More interestingly, the existence of nonzero $\Sigma_{0}$ in the obtained self-energy functional form is reminiscent of what is reported in multichannel Kondo problem \cite{Ludwig1991}, thereby suggesting that some novel quasiparticle behaviors
	and unconventional transport properties should be observed even in weakly disordered graphene.

	In the following, we digest the central findings in several important physical aspects.
	First, we discuss the relationship between the fitting parameters and disorder strength.
	In Fig.~\ref{fig:fitting}(a), a linear fitting of $\beta$ versus $\alpha\in[0.04,0.12]$ gives a slope of $2.00\pm0.05$. 
	$\Delta$ also has a linear relation with $\alpha$, and the fitting slope is $1.70\pm0.05$, as shown in Fig.~\ref{fig:fitting}(b). On the other hand, $\Sigma_{0}$ can be fitted by an exponential function $Ae^{-B/\alpha}$, and the fitting parameters are $A=1.0\pm0.1$ in unit of $t$ and $B=0.57\pm0.05$, as shown in Fig. \ref{fig:fitting}(c). Note that the exponential fitting
	parameter for $\Sigma_{0}$ is $B=0.57\pm0.05$, which is roughly a factor of 2 off the prediction (B=1) within the self-consistent Born approximation (SCBA) \cite{Ando,Ostrovsky}.

	\begin{figure}[t!]
		\includegraphics[width=8.5cm]{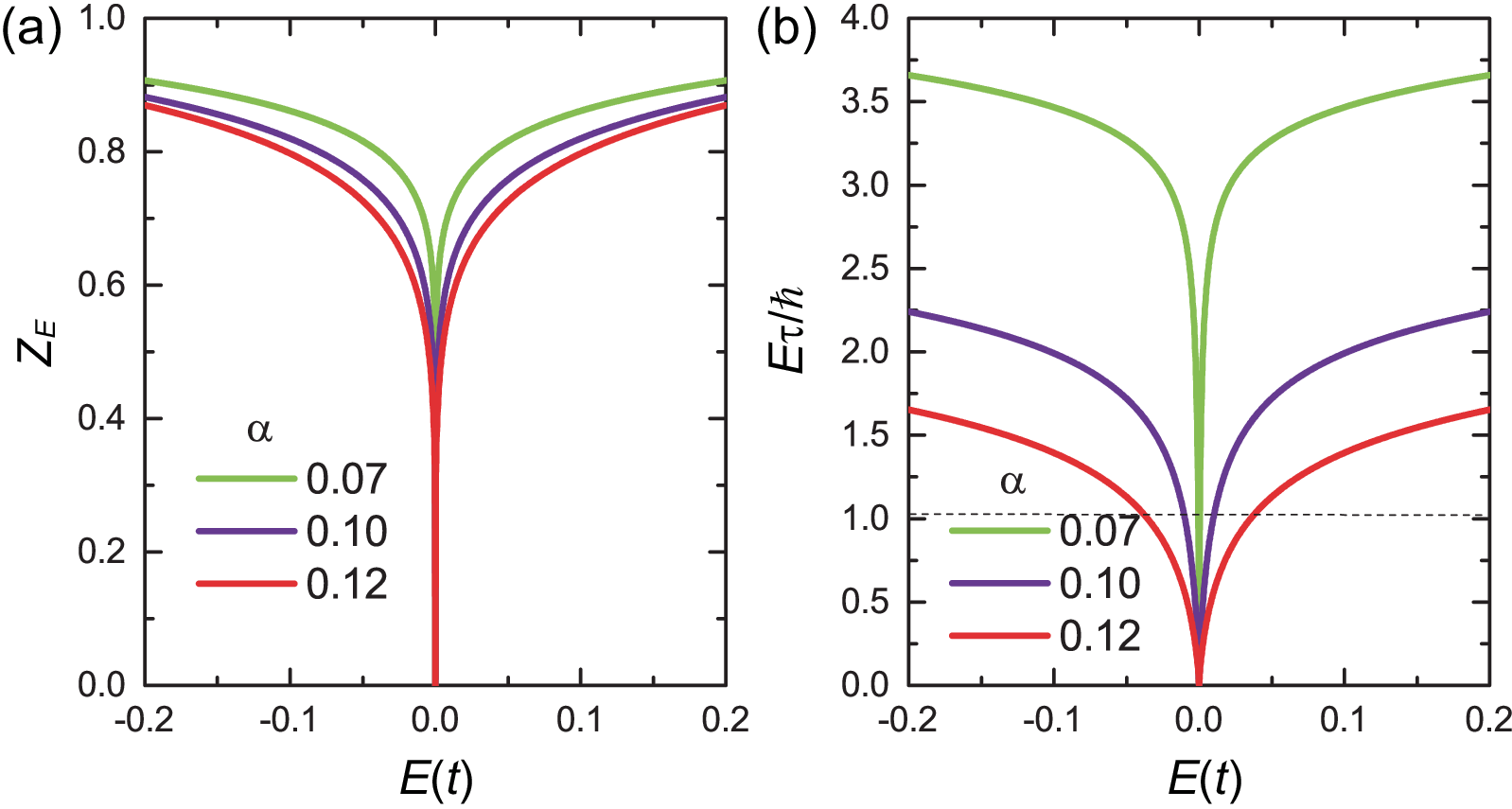} \caption{(a) Quasiparticle residue $Z_{E}=v_{g}/v_{f}$; (b) dimensionless parameter $E\tau/\hbar$ as a function of $E$ for different $\alpha$.}
		\label{fig.QP}
	\end{figure}
	
	Since all information of the quasiparticle properties is encoded in the self-energy function,
	next we discuss how the multi-scattering events considerably affect the quasiparticle
	properties. The real part of the self-energy ($\mathrm{Re}\Sigma$) in Eq.(\ref{Eq:sigma1}) contains
	two terms, a linear term $(CE)$ and a singular one $(D\mathrm{sgn}(E)|E|^{1-\beta})$.
	We find that the singular term will dominate the quasiparticle behavior around the Dirac point, leading to a super-linear dispersion of $E_{k}\propto k^{1/(1-\beta)}$, where $E_{k}$ is the root of $E-\hbar v_{f}k-\mathrm{Re}\Sigma(E)=0$. This result clearly indicates that the linear dispersion for the ideal graphene is unstable against disorder due to multiple scattering events. Moreover, the power-law correction for the real part of the self-energy leads to the quasiparticle residue $Z_{E} =1/[1-\partial_{E}\mathrm{Re} \Sigma(E)]\propto E^{\beta}$ vanishing as $E\rightarrow0$, so does the effective group velocity $v_{g}=\partial E_k/ \hbar \partial k=Z_{E}v_{f}$, as shown in Fig. \ref{fig.QP}(a). In the weak scattering limit ($E_{f}\tau/\hbar\gg1$), $Z_{E}$ is close to 1.0, and decreases slowly as the Fermi energy decreases. In the strong scattering limit, however, $Z_{E}$ (or $v_{g}$) drops rapidly to zero at the Dirac point. This unusual feature directly demonstrates that multiple scattering events significantly modify the quasiparticle properties near the Dirac point. Therefore, it is naturally expected that unconventional low-energy transport behaviors may arise in disordered graphene.
	\begin{figure}[t!]
		\includegraphics[width=8.5cm]{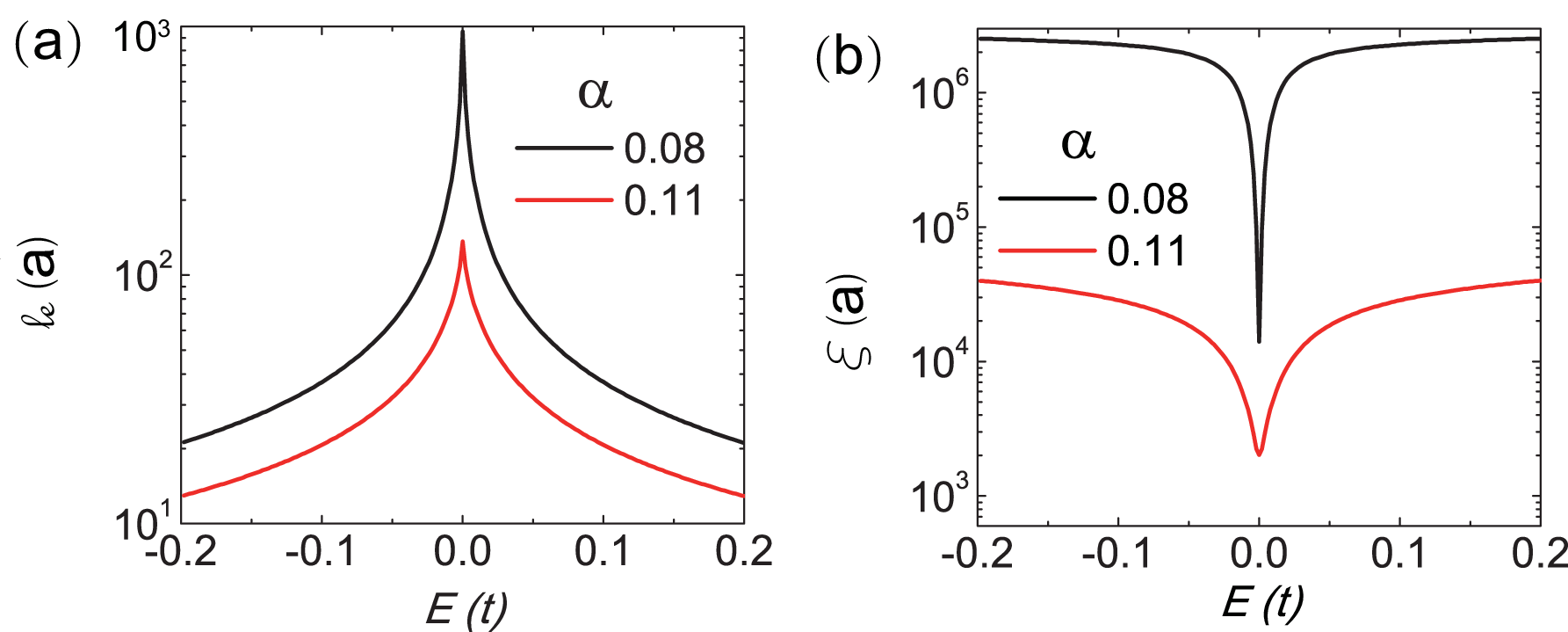} \caption{(a) The mean free path $\ell_{e}$ and (b) the localization length $\xi$ as a function of the energy for disorder strengths $\alpha=0.08$ and $0.11$.}
		\label{fig.lexi}
	\end{figure}
	
	Indeed, the elastic mean free path $\ell_{e}$ is given by the self-energy as $\ell_{e}=v_{g}\tau =3at/[-4\mathrm{Im} \Sigma(E)]$,
	where the elastic mean free time $\tau$ can be expressed as $\tau=\hbar/[-2Z_{E}\mathrm{Im}\Sigma(E)]$.
	Using our finding for $\Sigma_{0}$, the mean free path remains finite $\ell_{e}(0)\sim a\exp(0.57/\alpha)$ at the Dirac point, which is consistent with the results from the one-loop RG calculations \cite{Ostrovsky,schuessler2009}. But the lifetime $\tau$ diverges as $\propto E^{-\beta}$ in the limit $E\rightarrow0$, in stark contrast with the Fermi's golden rule prediction ($\tau\propto E^{-1}$), clarifying the significance of the multi-scattering events again. Fig.~\ref{fig.QP}(b) plots the dimensionless parameter ($E\tau/\hbar$) as a function of energy ($E$) in order to point out the low-energy window $|E|\approx E_{c}\exp (-1/2\alpha)$, which corresponds to the strong scattering regime ($E\tau/\hbar\le1$). As shown in Fig.~\ref{fig.lexi}(a), we plot the energy dependence of $\ell_{e}$ for $\alpha=0.08$ and $0.11$. With the decreasing disorder strength, $\ell_{e}$ at the Dirac point becomes longer. According to the scaling theory of Anderson localization, the 2D localization length ($\xi$) can be evaluated exclusively based on the diffusive transport properties $\xi=2\ell_{e}exp(\pi\sigma_{d}/G_{0})$ (orthogonal symmetry) \cite{ lee1985disordered,Fan14prb,Fan21review}, with $\sigma_d$ the conductivity of the system and $G_0=2e^2/h$. Fig. 3(d) shows that $\xi$ depends sensitively on the disorder strength and is strongly suppressed as $\alpha$ increases. The energy dependence of $\xi$ is mainly dominated by $\sigma_{d}$. As a result, the behavior of $\xi$ shows a minimum value at the Dirac point, exhibiting an opposite trend as $\ell_{e}$. Moreover, the localization length estimated by our numerical results agrees well with that obtained by the transfer matrix method \cite{Fan14prb}.

	\section{Transport behavior}
	\label{sec.transport}
	
	Based on the above self-energy results, we further investigate the transport properties of disordered graphene. First, we study the conductivity with impurity scattering including the multi-scattering events, and then take account the effect of electron-phonon scattering. At last, we consider the higher order correction in addition to the bare current bubble.
	
	\subsection{Conductivity With Impurity Scattering}
	\label{subsec.conductivity}
	
	\begin{figure}[t!]
		\includegraphics[width=8.5cm]{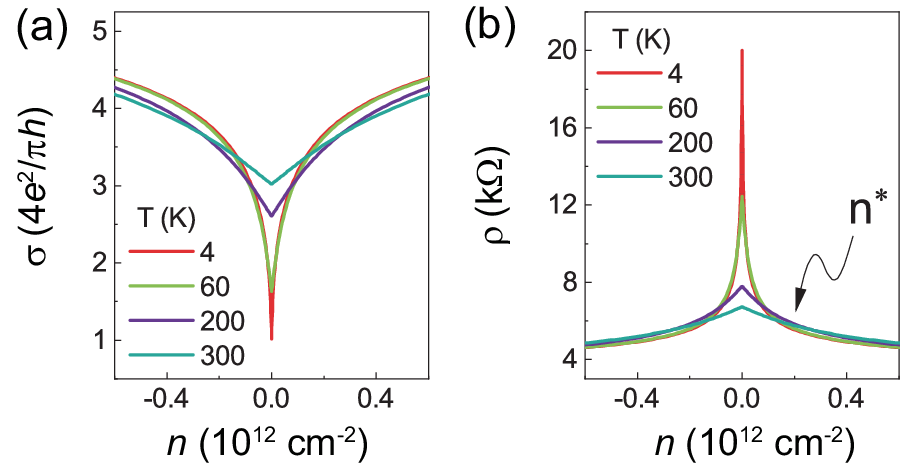} 
		\caption{(a) Conductivity $\sigma_{xx}$ and (b) resistivity $\rho$ as functions of the carrier density $n$ at different temperatures $T=\SI{4}{K}$, $ \SI{60}{K} $, $ \SI{200}{K} $, and $ \SI{300}{K} $. The disorder strength is $\alpha=0.09$. }
		\label{fig:cond}
	\end{figure}

On the Drude formula level, the conductance is given by $\sigma=\frac{e^{2}}{h}g$
with the dimensionless electrical conductance $g\sim k_{F}\ell_{e}$, where $k_{F}$
is the Fermi wave vector and $\ell_{e}$ is the mean free path. Using the Fermi energy
$E\sim \hbar v_f k_{F}$ and mean free path $\ell_{e}\sim v_{f}\tau$, the dimensionless
conductance can be rewritten as $g\sim E\tau/\hbar$. The conductance $g$
is a good measure of disorder and can be used as a parameter to interpolate between
the weak scattering regime $g\gg1$ and the strong scattering regime $g\le1$. Previous
theoretical studies are mainly restricted to extrinsic or doped graphene wherein the
Fermi level is away from the charge neutral Dirac point (or weak scattering regime).
The numerically exact results about the self-energies allow us to explore the transport
behaviors around the charge neutrality point where the dimensionless conductance
is not much larger than $1$. To include the non-trivial contribution of the quasiparticle residue, we take more rigorous quantum-mechanical treatments based on the Kubo formalism to calculate the Drude conductivity by	
	\begin{equation}
		\begin{aligned}\sigma_{xx}(T,E_{f}) & =\int dE\left(-\frac{\partial f(E,E_{f})}{\partial E}\right)\sigma_{d}(E),\end{aligned}
		\label{Eq:Zerow}
	\end{equation}
	\noindent where $f(E,E_{f})$$=1/[e^{(E-E_{f})/T}+1]$ is the Fermi-Dirac distribution
	with $T$ being the temperature, and $\sigma_{d}(E)$ is the zero temperature conductivity given as
%	\begin{equation}
%		\begin{aligned}\sigma_{d}(E)=\frac{G_{0}}{\pi}\left[1+\left(\chi(E)+\frac{1}{\chi(E)}\right)\tan^{-1}\chi(E)\right].\end{aligned}
%		\label{Eq:Drude}
%	\end{equation}
	\begin{equation}
			\begin{aligned}\sigma_{d}(E)=\frac{G_{0}}{\pi}\left[1+\chi(E)\tan^{-1}\chi(E) +\frac{\tan^{-1}\chi(E)}{\chi(E)} \right],\end{aligned}
			\label{Eq:Drude}
	\end{equation}
	with $G_{0}=2e^{2}/h$ and $\chi(E)=[E-\mathrm{Re} \Sigma(E)]/\mathrm{Im}\Sigma(E)$.
	After introducing a dimensionless function $\mathcal{G}(E)=\int_{0}^{E}\frac{Z(E)}{Z(E^{\prime})}\frac{dE^{\prime}}{E}=\frac{1-\mathrm{Re}\Sigma(E)/E}{1-\partial_{E}\mathrm{Re}\Sigma(E)}$, $\chi(E)$ can be rewritten as $\chi(E)=\mathcal{G}(E)E\tau/\hbar$. For a small disorder strength ($\alpha$ or $\beta \sim0$), our numerical calculation shows that $\chi(E)$ can be approximated as $E\tau/\hbar$. Thus, the Drude conductivity is only determined by the dimensionless parameter $E\tau/\hbar$. The conductivity (Eq.~(\ref{Eq:Drude})) contains two types of contributions: the first term (unity) in the bracket is the contributions of two Green's functions of the same kind (retarded-retarded or advanced-advanced) whereas the second and third terms come from the contribution of the retarded-advanced sector. In the weak scattering regime ($E\tau/\hbar\gg1$), the conductivity is dominated by the retarded-advanced term and takes the form $\sigma_{d}(E)\simeq \frac{G_{0}}{2} |\chi(E)|$, suggesting that weak disorder leads to weak dependence of conductivity on the Fermi energy. Around the Dirac point, however, the sublinear behavior of $E\tau/\hbar$ as plotted in Fig. \ref{fig.QP}(b) yields a sublinear power-law energy dependence of the obtained zero-temperature conductivity, in agreement with numerical calculations using the finite-size Kubo formalism \cite{Nomura2007}, but in sharp contrast with the prediction calculated by the Fermi's golden rule \cite{Hu}. More remarkably, it naturally produces the sharp peak in resistivity at low temperature and the strong temperature dependence of the maximum resistivity, due to the sharp dip of $E\tau/\hbar$ around the Dirac point as shown in Fig.~\ref{fig.QP}(b).
	Those novel behaviors have been widely reported in ultrahigh-mobility samples at and near the Dirac point \cite{Bolotin,Du,Ponomarenko,Morozov,Dean,Zomer,Mayorov2012,Nuno,kazi2018,WangLujun}.
	
	To compare with the experimental transport results of high quality graphene in more detail, in the following quantitative evaluations,
	a typical weak disorder strength is chosen as $\alpha=0.09$
	without any other adjustable parameter being used. 
	Fig.~\ref{fig:cond}(a) and ~\ref{fig:cond}(b) plot the corresponding conductivity $\sigma_{xx}$ and resistivity
	$\rho=1/\sigma_{xx}$ as functions of the carrier density $n$ from the temperature $ \SI{4}{K} $ to $ \SI{300}{K} $, respectively,
	where $ n=\int_0^{\infty} D(E) f(E,E_f)dE +\int_{-\infty}^0 D(E) [1-f(E,E_f)]dE $, and $D(E)$ denotes the density of states. Sharp dips (or
	peaks) in the conductivity (or resistivity) are observed precisely at the Dirac point
	at low temperatures. By increasing the temperature, the conductivity very close to
	the Dirac point has a pronounced increase, showing a strong temperature dependence.
	Most remarkably, there exists a $T$-independent carrier density (roughly $ n^{\star}\sim \SI{1.5e11}{cm^{-2}} $)
	that divides the systems into two different density regimes. In the low density regime
	($|n|<n^{\star}$), the resistivity exhibits a nonmetallic behavior, that is, increasing
	$\rho$ for decreasing $T$. For $|n|>n^{\star}$, the resistivity displays a weak
	$T$-dependence and decreases for decreasing $T$.

	\begin{figure}[t!]
		\includegraphics[width=8.5cm]{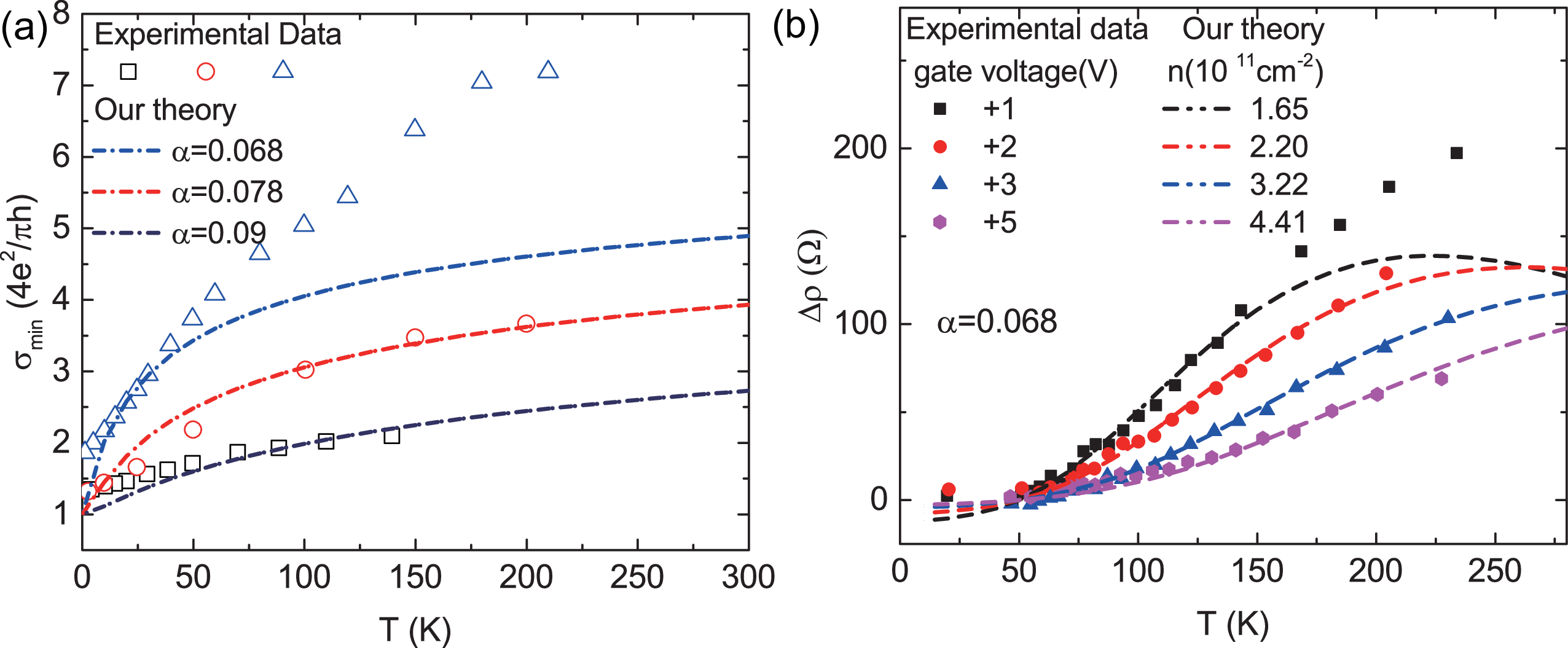} \caption{Comparisons between our theory and the experimental data for the (a) minimum conductivity $\sigma_{\text{min}}$ and (b) resistivity $\Delta \rho$ as function of temperature. In (a), the open symbols indicate the temperature dependence of $\sigma_{min}$ at the Dirac point for three devices extracted from Ref. \cite{Mayorov2012}. The dashed lines indicates the results according to our theory for three different disorder strengths $\alpha$. In (b), the solid symbols represent the temperature dependence of $\Delta \rho$ for different gate voltages extracted from Ref. \cite{Bolotin}. The dash lines are the results according to our theory for different carrier densities.}
		\label{fig:comparison}
	\end{figure}

We separately consider the two density regimes $|n| < n^{\star}$
and $|n| > n^{\star}$ and compare our theory with experimental results. We first consider low density regime, $|n| < n^{\star}$, and
address the $T$ dependence of the minimum conductivity. Fig.~\ref{fig:comparison}(a) shows the comparison of the minimum conductivity $\sigma_{\text{min}}$  as a function of temperature between our theory and experimental data for three monolayer devices from Ref. \cite{Mayorov2012}. According to our theory, $\sigma_{\text{min}}$ increases monotonically with $T$. $\sigma_{\text{min}}$ versus $T$ follows a roughly linear relationship for $T<\SI{100}{K}$ and becomes sublinear for $T>\SI{100}{K}$. The experiment and theory show good agreement for devices $\square$ and $\bigcirc$. For device $\triangle$, the theory only fits the experimental data well at low temperature.  At finite temperature, electrons in both the conduction band and the valence band can contribute to the electrical conductivity. From Eq.(\ref{Eq:Zerow}),  the broadening width of the electron and hole contributions is proportional to the temperature according to the Fermi-Dirac distribution $f(E,E_f)$. As the temperature increases, the broadening width also increases, allowing more electron-hole pairs to contribute to the electrical conductivity. The temperature dependence of $\sigma_{\text{min}}$ depends critically on the transport properties near the Dirac point. 
We then turn to the high density $|n|>n^{\star}$. We depict $\Delta\rho(T)=\rho(T)-\rho(\SI{50}{K})$ as a function of temperature with different carrier densities in Fig.~\ref{fig:comparison}(b). The solid dots and dashed lines are experimental data from Ref. \cite{Bolotin} and our theory, respectively. In the high temperature range, $\Delta\rho(T)$ increases nearly linearly with $T$. In Ref. \cite{Bolotin}, the linear temperature dependence is believed to be due to electron-phonon interaction. However, the slope of $\Delta\rho$ versus $T$ cannot be explained solely by electron-phonon interaction, as it also depends on the carrier density. Our theory can consistently explain the carrier density and temperature dependence of $\rho(T)$.  The overall trends of our numerical results are in good agreement with experimental observations \cite{Bolotin}. The discrepancy at high temperatures could be due to the neglect of electron-electron scattering and electron-phonon scattering, which become significant at high temperatures. Those findings attest that the strong $T$-dependence of the conductivity (resistivity) in the low density regime stems from the multi-scattering effects, which amounts to another important aspect of the present work.
	
	\begin{figure}[t!]
		\includegraphics[width=8.5cm]{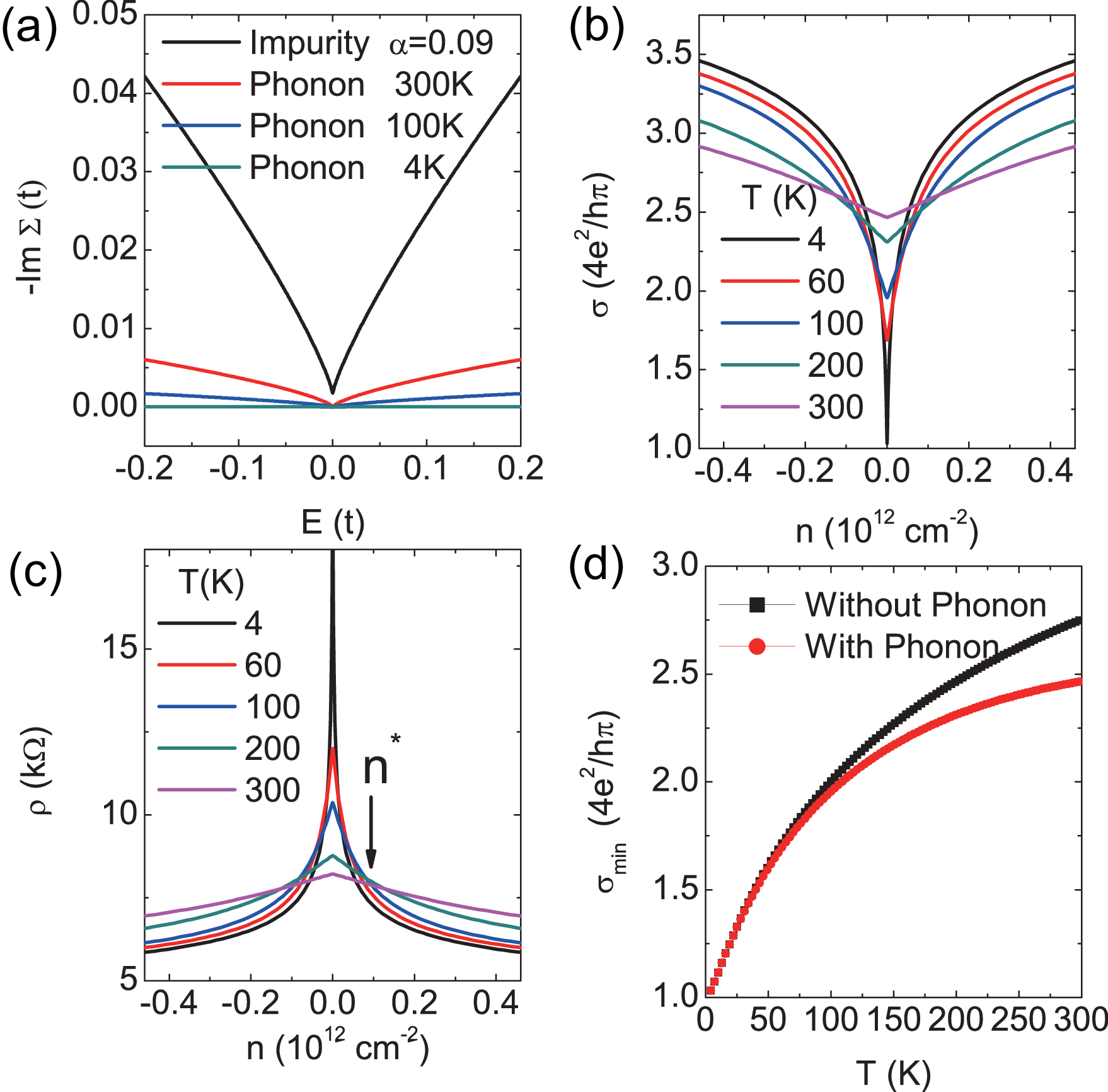} \caption{(a) Imaginary part of the self-energy induced by disorder scattering and phonon scattering at $\SI{4}{K}$, $\SI{100}{K}$, and $\SI{300}{K}$.(b) Conductivity $\sigma_{xx}$ and (c) resistivity $\rho$ as functions of the carrier density $n$ at different temperatures $T=\SI{4}{K}$, $\SI{60}{K}$, $\SI{100}{K}$, $\SI{200}{K}$ and $\SI{300}{K}$ including resistive scattering by graphene phonons described by Eq. (\ref{eq.phononscattering}). The disorder strength is $\alpha=0.09$. (d) Comparison of temperature dependence of the minimum conductivity $\sigma_{\text{min}}$ with and without phonon scattering. }
		\label{fig:phonon}
	\end{figure}

	\subsection{Conductivity With Phonon Scattering}
	\label{subsec.phonon}
	We are now going to take into account the effect of electron-phonon scattering. In
	graphene, there exists a characteristic wave vector $q_{c}$ below which the anharmonic
	effects become important \cite{Gornyi,EVCastro}. It has be estimated that $q_{c}=\sqrt{\Delta_{c}T}/(\hbar v_{f})\approx\sqrt{T(K)}\times \SI{0.7e8}{m^{-1}} $,
	where $\Delta_{c}\approx \SI{18.7}{eV} $ \cite{Gornyi}. Since our interest is the low carrier
	density with $ n\le \SI{0.5e12}{cm^{-2}} $ where the transport properties are
	strongly influenced by the multiple scattering processes. The Fermi wave vector can
	be estimated by $k_{F}\approx\sqrt{\pi n}\le \SI{1.25e8}{m^{-1}}$, and
	is small compared to $q_{c}$ ($\SI{3.2}{K}$). Therefore, the anharmonic electron-phonon interaction
	should be taken into account. In this situation, the scattering rate caused by phonon
	scattering can be expressed as \cite{Gornyi}
	\begin{align}
		\frac{\hbar}{2\tau(\epsilon)}=\frac{g^{2}T^{2}}{\pi|\epsilon|}CZ^{2}(\frac{|\epsilon|}{\sqrt{T\Delta_{c}}})^{2\eta},
		\label{eq.phononscattering}
	\end{align}
	where $g\approx5.3$ is the dimensionless electron-phonon coupling constant, $C\approx2.26$
	is an integral coefficient, $\eta\approx0.85$ is a critical index \cite{Kownacki}, and
	the numerical prefactor $Z\approx1$.
	
	As shown in Fig.~\ref{fig:phonon}(a), we compare the magnitude of our calculated imaginary part of the self-energy $\mathrm{Im}\Sigma$ due to impurity scattering and the contribution arising from the electron-phonon scattering (Eq. (\ref{eq.phononscattering})) at different temperatures
	$T=4,100,\SI{300}{K}$. The disorder strength has been chosen as $\alpha=0.09$. At low temperatures $(T<\SI{100}{K})$, $-\mathrm{Im}\Sigma\gg\frac{\hbar}{2\tau(\epsilon)}$,
	and the resistivity of graphene is dominated by scattering of impurities. We also plot the resistivity and conductivity after taking phonon scattering into consideration in Figs.~\ref{fig:phonon}(b) and ~\ref{fig:phonon}(c), and the temperature dependences of the minimum conductivity $\sigma_{\text{min}}$ with and without phonon scattering are contrasted in Fig.~\ref{fig:phonon}(d). By comparing with the results in sec.~\ref{subsec.conductivity}, we find that our main conclusion would not change even considering the electron-phonon scattering. The crossover carrier density $n^{\star}$ still exists, separating the regions with the "metallic" (high density) and "insulating" (low density) behaviors.
	
	\subsection{Higher Order Conductivity Correction}
	In addition to the bare (zeroth order) current bubble which yields the main contribution to the classical conductivity,
	the disorder averaging will generate other current bubbles which are expanded in terms of scattering vertices. Two classes of diagrams are usually calculated, the ladder diagram and the maximally-crossed diagrams, which account for the vertex correction and quantum interference correction, respectively. 
	
	\subsubsection{Vertex Correction}
	The Bethe-Salpeter Fermi equations for the vertex correction can be solved by using the single-particle propagators with the full self-energy. With the vertex correction, the Kubo formula for conductivity is given by
	\begin{align}
		\begin{split}
			\sigma_{xx}(E)= & -\frac{\hbar e^{2}v_{f}^{2}}{4\pi}\sum_{ss^{'}=\pm}ss^{'}\int\frac{d^{2}\boldsymbol{k}}{(2\pi)^{2}}\mathrm{Tr}[j_{x}G(\boldsymbol{k},a+is\eta)\\
			& \times J_{x}(\boldsymbol{k},a+is\eta,a+is^{'}\eta)G(\boldsymbol{k},a+is^{'}\eta)].
		\end{split}
		\label{kubo}
	\end{align}
	Here the current vertex $J_{x}$ satisfies the following Beta-Salpeter equation \cite{Shon}:
	\begin{equation}
		\begin{aligned}  &J_{x}(\boldsymbol{k},a+is\eta,a+is^{'}\eta) =j_{x}+\sum_{\boldsymbol{k}^{'}} \langle V_{\boldsymbol{k}-\boldsymbol{k}^{'}} G(\boldsymbol{k}^{'},a+is\eta)
		\\ & \times J_{x}(\boldsymbol{k}^{'},a+is\eta,a+is^{'}\eta)
		\times G(\boldsymbol{k}^{'},E+is^{'}\eta) V_{\boldsymbol{k}^{\prime}-\boldsymbol{k}}\rangle_{\text{dis}},
		\end{aligned}
		\label{vertex-1}
	\end{equation}
	where we have defined $E-\mathrm{Re}\Sigma\equiv a$ and $-\mathrm{Im}\Sigma\equiv\eta$
	for simplicity and $G(\boldsymbol{k},a+is\eta) =1/(a+is\eta-\hbar v_{f}\boldsymbol{k}\cdot \boldsymbol{\sigma})$ is the disorder averaged retarded $(s=+)$ and advanced $(s=-)$ Green's functions with our calculated
	self-energy. By further assuming that compared with the bare current $j_{x}=\sigma_{x}$, the renormalized current $J_{x}=\Lambda\sigma_{x}$ only differs by an energy 
	%and disorder strength 
	dependent dimensionless coefficient $\Lambda$, we can put it into the iterative equation of Eq.~(\ref{vertex-1}).
	% After considering the vertex correction, the conductivity will be renormalized by the dimensionless coefficient $\Lambda$ which depends on disorder strength and the Fermi energy.
	
	For the short range disorder, after taking inter-valley scattering into account, the vertex correction can be shown to vanish identically due to the symmetry of the first Brillouin zone. 
	Therefore, the vertex correction only contributes in the long range disorder case. As shown in Appendix.~\ref{subsec.vertex}, with the vertex correction, the minimum conductivity will be dependent on the disorder strength.
	For the Fermi energies far from the Dirac point, the vertex correction $\Lambda=2$ recovers the result for the weak scattering regime \cite{Ando,McCann}. 
	As one gets close to the Dirac point, the vertex correction becomes negligible due to the sharp reduction near the Dirac point and eventual vanishment at the Dirac point of the quasiparticle residue as plotted in Fig.~\ref{fig.QP}(a).

	\subsubsection{Quantum Interference Corrections}
	Another mystery in graphene transport is the absence of the localization-induced insulating phase in the vicinity of the Dirac point, violating the Ioffe-Regel criterion which states that the electron state will be localized in the region $E\tau/\hbar \ll1$ \cite{Ioffe,lee1985disordered}. In undoped samples of graphene, the minimum conductivity is observed to remain almost constant over a wide range of temperatures, from room temperature down to sub-Kelvin temperatures\cite{Du,Mayorov2012}. This behavior is in stark contrast to the well-established results on the conductivity of 2D systems, where localization effects typically drive the system into an insulating state at low temperatures. The absence of localization in graphene is still not fully understood.
	Our calculations show that the multi-scattering events may provide a plausible mechanism in understanding the absence of the localization. In realistic graphene samples, inter-valley scattering is inevitable, leading to backscattering between the two valleys. As a result, the inter-valley Cooperon channel dominates at small magnetic fields or large sample sizes, leading to weak localization effects and even localization when the quantum interference correction becomes comparable to the classical conductivity. This is the reason why the magnetoresistance in experiments at small magnetic fields is commonly negative, exhibiting a weak localization behavior \cite{Morozov2006prl,Tikhonenko2009prl}.
	Here, we extend the standard calculation from the weak scattering limit \cite{McCann,Fu2019prl} to the strong
	scattering limit to discuss the contribution of the maximally-crossed diagrams by considering the accurate single-particle propagator, shown in Appendix.~\ref{subsec.QI}. In the weak scattering regime, we recover the weak localization correction which arises from the inter-valley
	scattering induced Cooperon channel \cite{McCann}. In the strong scattering regime, however, we verify that multi-scattering events will introduce finite Cooperon gaps so that the small momentum singularities in the Cooperon momentum integrals is avoided. Thus, the weak localization correction is strongly suppressed in the vicinity of the Dirac point. This may explain why the Anderson localization is absent in the transport measurements in graphene  \cite{Du,Ponomarenko,Mayorov2012}.
	
	\section{Correlated Impurities}
	\label{sec.corrlated potential}
	
	\begin{figure}[t!]
		\includegraphics[width=8.5cm]{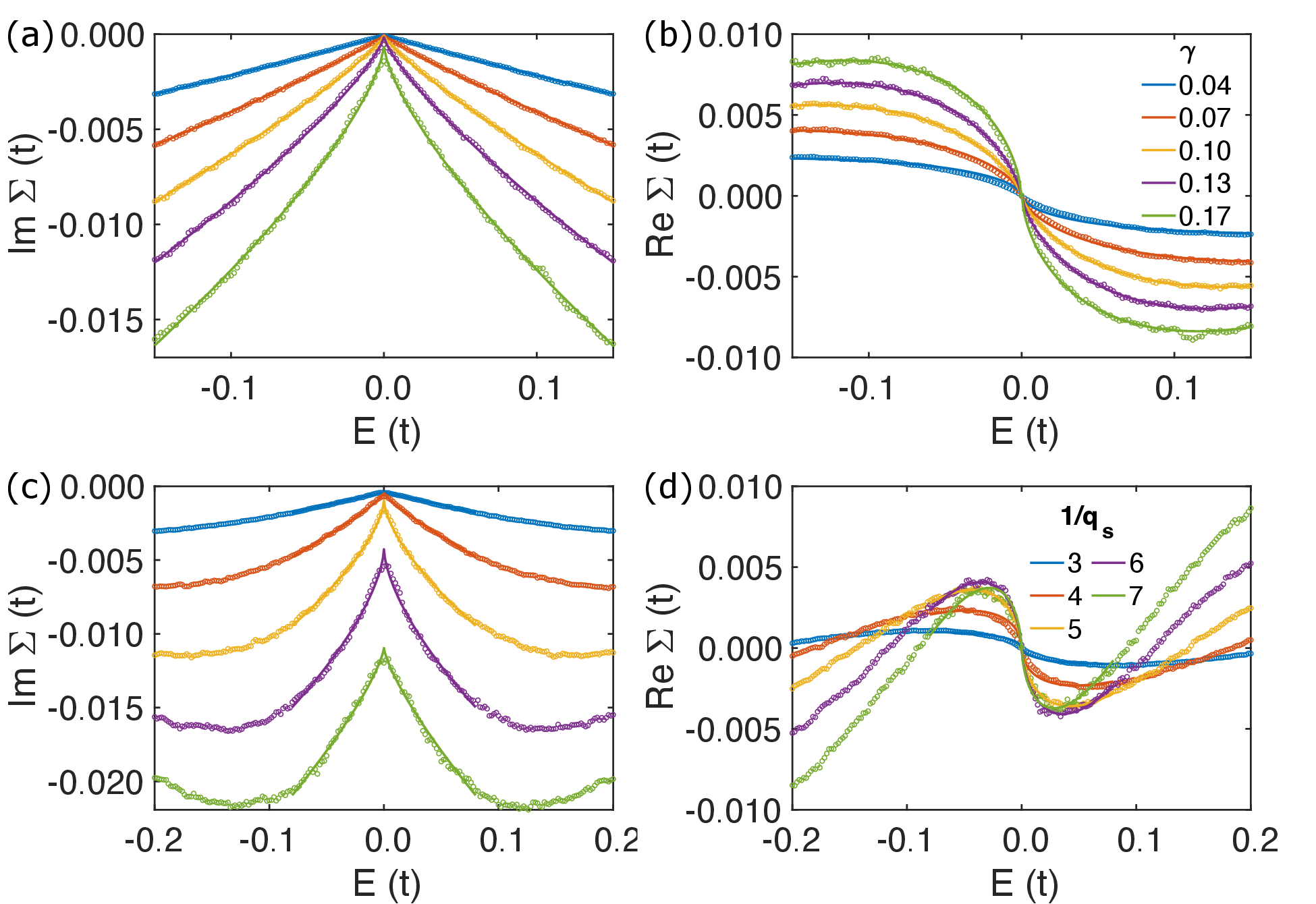} \caption{(a) Imaginary and (b) real parts of the self-energy with a correlated Gaussian potential as a function of energy for different disorder strengths $(0.04\leqslant\gamma\leqslant0.17)$.
		The circle symbols denote the numerical results obtained from our Lanczos method and the solid lines are the fitting curves. $60$ samples are collected for each curve. (c) Imaginary and (d) real parts of the self-energy with Yukawa-type charge impurities as a function of energy for different screening lengths $ 1/q_s =3,4,5,6,7 $ ($ \gamma_c=0.04,0.11, 0.22,0.39, 0.61 $). The circle symbols denote the numerical results obtained from the Lanczos method and the solid lines are the fitting curves within the energy window [-0.08t,0.08t].}
		\label{fig:self_corr}
	\end{figure}

 To simulate various defects in real experimental conditions, we expand our regime of discussions to the cases where each impurity has a finite range. 
 In such cases,  two impurities become correlated, and the systems can be characterized as containing correlated potential disorder. 
 Since the correlated potential disorder is smooth at the atomic scale, the inter-valley scattering or backscattering is suppressed. 
 For a correlated potential, the self-energy depends on the energy $E$ and wave vector $\boldsymbol{k}$, which can be directly obtained by changing the initial state $|\boldsymbol{k}\rangle$ in our numerical method \cite{zhu2010}. Here we only focus on the self-energy $\Sigma(E)$ for $k=0$, which is symmetric about $E=0$. 
 Note that the wavelength of the low-energy quasiparticle approaches infinite in the vicinity of the Dirac point. 
 Therefore, a random potential with a shorter spatial correlated length cannot be seen by the Dirac electronic wave, and it will not influence qualitatively the quasiparticle (self-energy) behaviors. Here we also use the power-law formula {[}see Eqs.~(\ref{Eq:imaginaryselfenergy}) and (\ref{Eq:sigma1}){]} to fit the numerical results for the correlated disorder potentials.

	\subsection{Gaussian Potential}
	First, we consider the most common type, Gaussian correlated disorder potential $V_{i}=\sum_{n=1}^{N_{\text{imp}}}\pm u_{0}\exp[-|\boldsymbol{r}_{n}-\boldsymbol{r}_{i}|^{2}/(2\xi^{2})]$, where $\xi$ is the Gaussian correlation
	length. The scatters of $\pm u_{0}$ are randomly distributed with equal probability,
	and $N_{\text{imp}}$ impurities are randomly located among the $N=4000^{2}$ lattices. We fix
	the impurity density $n_{\text{imp}}=N_{\text{imp}}/N=1\%$ and take $\xi=2a$ as an example in the following
	calculations. After the disorder averages, the disorder potential has a vanishing mean and a smooth form of the correlator:
	\begin{align}
		\langle V_{i}\rangle_{\text{dis}} & =0,\\
		\langle V_{i}V_{j}\rangle_{\text{dis}} & =\gamma\frac{(\hbar v_{f})^{2}}{4\pi\xi^{2}}e^{-|\boldsymbol{r}_{i}-\boldsymbol{r}_{j}|^{2}/4\xi^{2}},
		\\ \langle V_{\boldsymbol{k} -\boldsymbol{k}^\prime} V_{\boldsymbol{k}^\prime -\boldsymbol{k}}  \rangle_{\text{dis}} &=\gamma (\hbar v_f)^2 e^{-\xi ^2 |\boldsymbol{k}- \boldsymbol{k}^\prime|^2 },
	\end{align}
	where $\gamma=\frac{n_{\text{imp}}u_{0}^{2}}{A_{c}}\frac{(2\pi\xi^{2})^{2}}{(\hbar v_{f})^{2}}$
	is the dimensionless disorder strength. As shown in Figs.~\ref{fig:self_corr}(a) and \ref{fig:self_corr}(b), the agreement between the power-law fitting and
	numerical results obtained for the Lanczos method is very good within the energy window of $[-0.15t,0.15t]$. 
	The linear fittings of $\beta$ and
	$\Delta$ with the disorder strength $\gamma$ are given in Appendix.~\ref{subsec.Gaussian}, yielding $\beta=2.04\gamma$ and $\Delta=0.42\gamma$.

\begin{figure}[t!]
		\includegraphics[width=8.5cm]{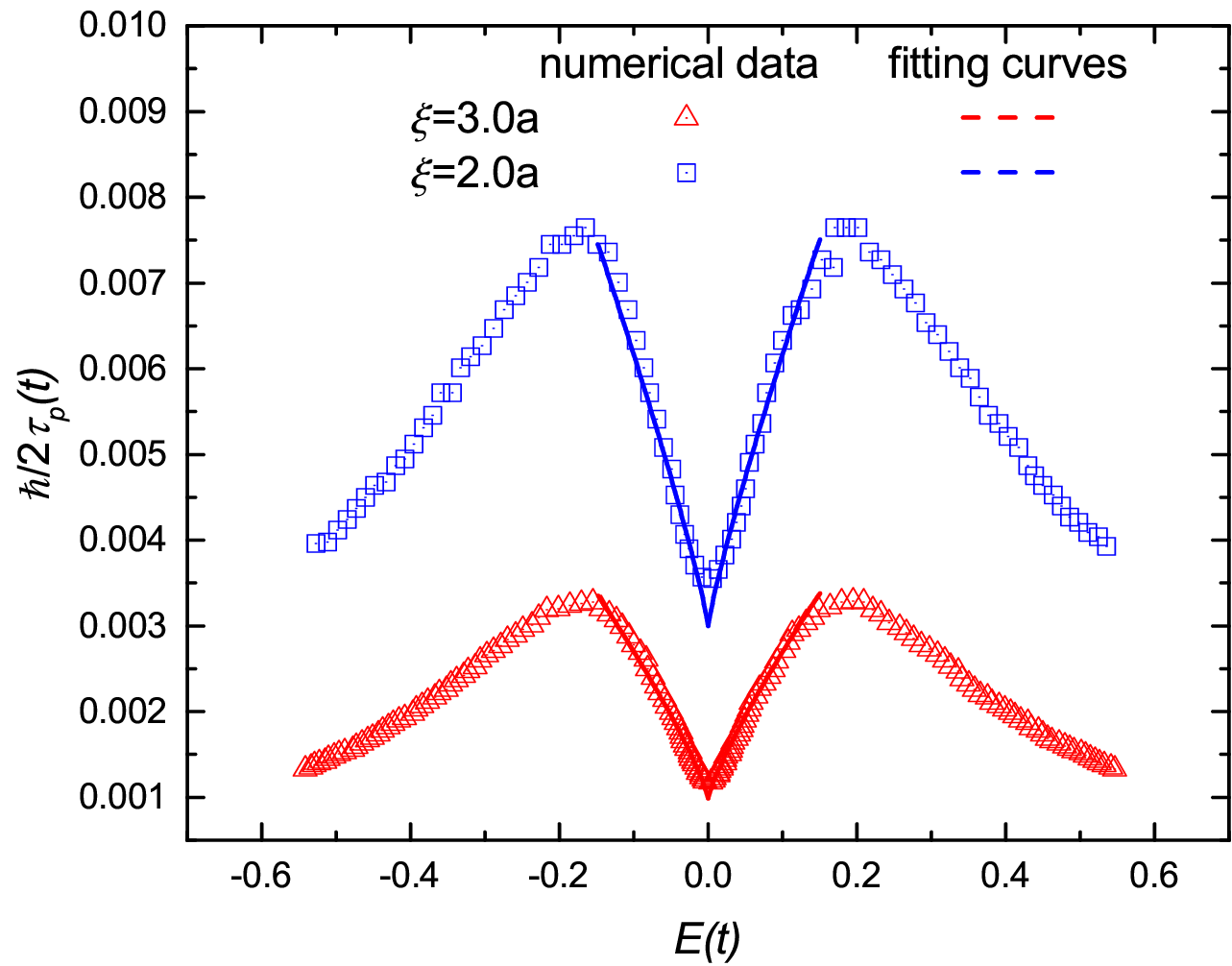} \caption{The fitting of the inverse of the energy-dependent momentum relaxation time $\tau_p  (E)$ by using the power-law formula. The open symbols denote the numerical results extracted from Fig. 16 in Ref.\cite{Fan21review} and the dashed lines are the fitting curves.}
		\label{fig:comparewithreal}
	\end{figure}

We compare the results of this manuscript with that in the existing literature based on real-space methods and refer to \cite{Fan21review}, where the momentum relaxation time $\tau_p(E)$ was investigated. Inverting $\tau_p (E)$ yields the imaginary part of the self-energy, which can be expressed as $-\mathrm{Im}\Sigma^R (E)=\hbar/[2\tau_p(E)]$. We then fit our proposed power-law formula to the data, and as shown in Fig.~\ref{fig:comparewithreal}, the formula is a good fit in the low energy regime for $E\in[-0.15t,0.15t]$ (equivalently $[\SI{-0.405}{eV},\SI{0.405}{eV}]$).

	\subsection{Yukawa Potential}

	Considering the observation of electron-hole puddles, we also consider the case of a Yukawa-type potential $ V_i =\sum_{n=1}^{N_{\text{imp}}} \pm \frac{e^2}{\kappa |\boldsymbol{r}_n -\boldsymbol{r}_i|} \exp[-q_s |\boldsymbol{r}_n -\boldsymbol{r}_i|] $, with positive and negative charged impurities possessing equal probabilities, where $ e $ is the electron charge, $ r_s=e^2/\hbar v_f=2.2 $ is a constant \cite{Reed}, $ \kappa $ is the background dielectric constant, and $ q_s $ is the inverse screening length. The charge impurities are randomly distributed in the substrate, and we fix the impurity concentration to be $ n_{\text{imp}}=0.25\%  $ ($ \sim \SI{e12}{cm^{-2}} $) and the distance between the charge and graphene plane to be $ d=3a $ in the following calculations. By the Fourier transformation, in momentum space, the potential is given by
	\begin{align}
		V_{q} &=  \frac{ V_0}{\sqrt{q_s^2+ q^2} } e^{-d\sqrt{q_s^2+ q^2 }}.
	\end{align}
	After averaging the disorder, we obtain
	\begin{align}
		\langle V_i \rangle_{\text{dis}}&=0,
		\\ \langle V_{\boldsymbol{k} -\boldsymbol{k}^\prime} V_{\boldsymbol{k}^\prime -\boldsymbol{k}}  \rangle_{\text{dis}}&= \frac{n_{\text{imp}}}{A_c} \frac{ V_0^2}{q_s^2+ |\boldsymbol{k}-\boldsymbol{k}^\prime|^2} e^{-2d\sqrt{q_s^2+ |\boldsymbol{k}-\boldsymbol{k}^\prime|^2 }},
	\end{align}
	where $ V_0=2\pi e^2/\kappa $. Here we define the dimensionless disorder strength as $ \gamma_c= \frac{n_{\text{imp}}}{A_c} \frac{ (2\pi r_s)^2}{\kappa^2 q_s^2} e^{-2q_s d} $. As shown in Figs.~\ref{fig:self_corr}(c) and \ref{fig:self_corr}(d), the self-energies are given for different screening lengths. In a small energy window, such as $ [-0.08t,0.08t] $, the power-law formula can be still used to fit the behavior of the self-energy. At higher energies, this formula does not work because the corresponding Fermi wave vector is larger and the electron wavelengths are comparable to the correlation lengths in this region. 

We then compare the results of this manuscript with the existing literature based on real-space methods.
	% When the disorder strength increases, the quasiparticle lifetime decreases and elastic mean free path $ \ell_{e}=3at/[-4\mathrm{Im} \Sigma(E)]  decreases. % The real-part of self-energy leads to the quasiparticle residue $Z_{E} =1/[1-\partial_{E}\mathrm{Re} \Sigma(E)) $ increasing as $ |E| $ increase and at some critic point, $ Z_E>1 $.  The corresponding energy range of the strong scattering region decreases gradually.
	
	\section{Conclusion}
	\label{sec.discussion}
	
	In summary, using the numerically exact momentum-space Lanczos method, we have systematically investigated the multiple impurity scattering effects on the quasiparticle and transport properties of two-dimensional Dirac fermionic systems in the presence of isolated or correlated weak scalar potentials. We uncover that the multiple impurity scattering processes arising from the weak disorder can induce nontrivial non-Fermi liquid behavior, which is insensitive to the detailed types of disorder. Our theory can account for a set of   unconventional findings in the transport measurements: (i) The temperature-dependent resistivity can be divided into two different density regimes: a metallic regime and an insulating regime, separated by $n^{\star}$. (ii) For $|n|<n^{\star}$, we examined the temperature dependence of the minimum conductivity at the Dirac point. As the temperature increases, the temperature-dependent minimum conductivity first increases linearly, then becomes sublinear, and tends to saturate at higher temperatures. (iii) In the high-density regime $|n|>n^{\star}$, the resistivity linearly increases with temperature in the high-temperature range when $n$ is not too close to $n^{\star}$. The slope of the resistivity versus temperature increases as $n$ gets closer to $n^{\star}$. Our theory can consistently explain the temperature and carrier density dependence of conductivity.  Our work attests that the vital importance of multiple impurity scattering events in understanding the exotic low energy physics of ultrahigh-mobility graphene. 
	
	\begin{acknowledgments}
		We thank Profs. Xin-Cheng Xie, Qing-Feng Sun, and Xiang-Rong Wang for valuable discussions.
		This work was supported by National Key Research $\&$ Development Program of China
		(No. 2016YFA0200600 and 2017YFA0204904), National Natural Science Foundation of China (No. 21473168, 11634011, 11774325, 12047544, 21603210 and 11974323), Fundamental Research Funds for the Central Universities and the Innovation Program for Quantum Science and Technology (Grant No. 2021ZD0302800). Computational resources are provided
		by CAS, Shanghai and USTC Supercomputer Centers.
	\end{acknowledgments}

	\appendix

	\section{Spectral Function}
	\label{subsec.SF}
\begin{figure}[t!]
		\includegraphics[width=8.5cm]{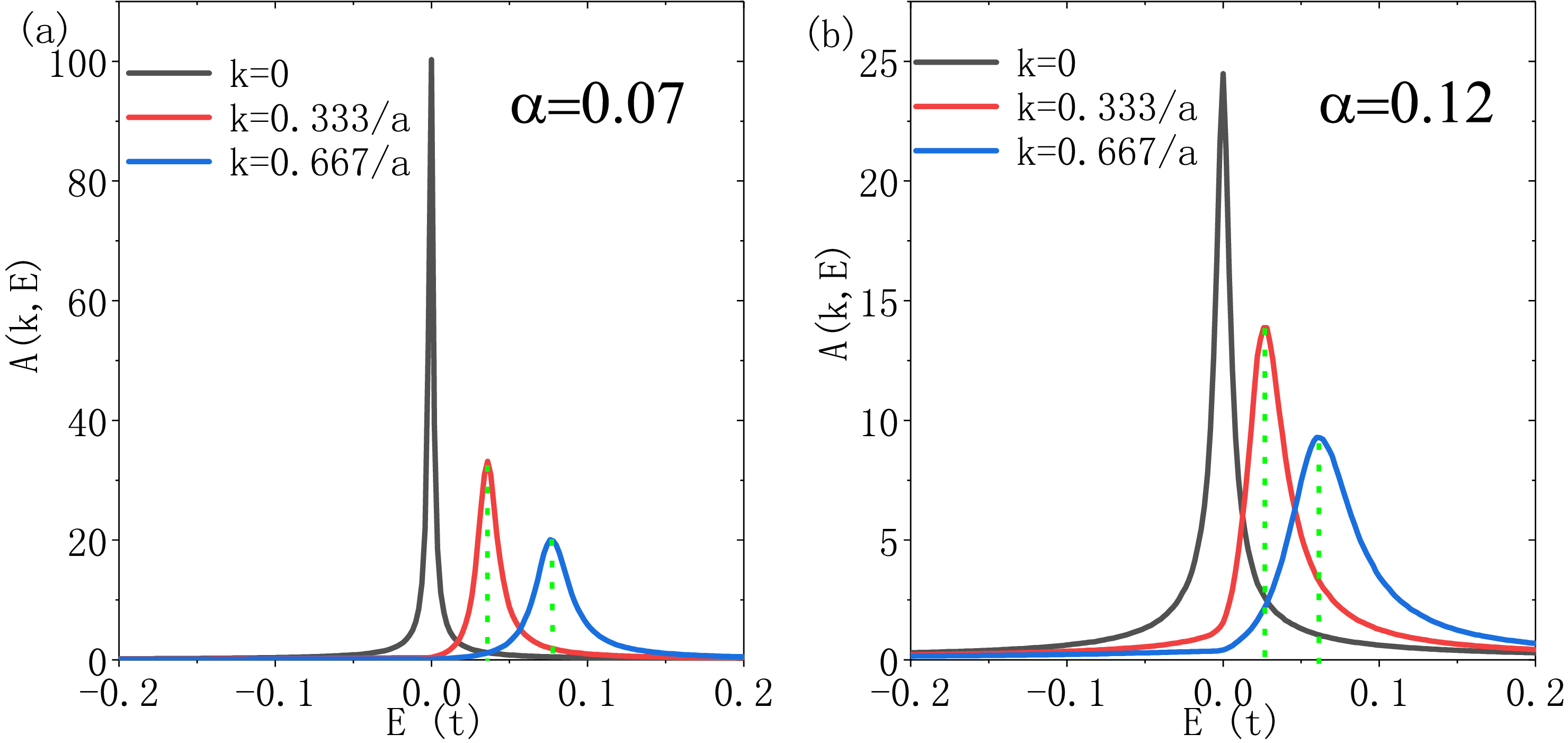} 
		\caption{Single-particle spectral function $A(\boldsymbol{k}+,E)$ (conduction band) plotted as a function of energy $E$ at several k points of $k=0.00$ (black lines), $0.333/a$ (red lines), $0.666/a$ (blue lines) along the $k_{x}$ direction (from left to right). The disorder strength is chosen to be (a) $\alpha=0.07$ and (b) $\alpha=0.12$, respectively.}
		\label{fig:spectral}
	\end{figure}

	%and blow this energy scale SCBA gives an unphysical result of the group velocity at the Dirac point($v_g(0)=v_f\alpha/(1+\alpha)$ which grows as the disorder increases).
	To demonstrate how this power-law correction significantly renormalizes the quasiparticle
	properties around the Dirac point more intuitively, we calculate the spectral
	function. The single-particle spectral function relates to the Green's function
	through
	
	\begin{equation}
		\begin{aligned}A(s\boldsymbol{k},E) & =-\frac{1}{\pi}\mathrm{Im}G(s\boldsymbol{k},E)\\
			& =\frac{1}{\pi}\frac{-\mathrm{Im}\Sigma}{(E-s\hbar v_{f}k-\mathrm{Re}\Sigma)^{2}+(\mathrm{Im}\Sigma)^{2}}\\
			& =\frac{1}{\pi}\frac{\eta}{(a-s\hbar v_{f}k)^{2}+\eta^{2}},
		\end{aligned}
		\label{eq.spectral}
	\end{equation}
	where $s=\pm$ represents the conduction band and valance band respectively, and
	we have defined $E-\mathrm{Re}\Sigma\equiv a$ and $-\mathrm{Im}\Sigma\equiv\eta$
	for simplicity. In the absence of disorder, the spectral function $A(\boldsymbol{k},E)$
	is a $\delta$ function, reflecting that the wave vector $k$ is a good quantum number
	and all its weight ratio is precisely at $E=s\hbar v_{f}k$. In the presence of disorder,
	Eq.~(\ref{eq.spectral}) is plotted graphically in Fig.~\ref{fig:spectral}. For $\alpha=0.07$, $A(\boldsymbol{k}=0,E)$
	exhibits a sharp peak of a Lorentzian type at $E=0$ shown as the black line in Fig.~\ref{fig:spectral}(a). When $k$ moves away from the Dirac point, $A(\boldsymbol{k},E)$ maintains the Lorentzian
	line shape but becomes much broader due to the increasing of the scattering processes
	(red and blue lines in Fig.~\ref{fig:spectral}(a)). For $\alpha=0.12$, $A(\boldsymbol{k},E)$ clearly
	deviates from the Lorentzian type and carries substantially more weight in wings as shown
	in Fig.~\ref{fig:spectral}(b). One can also extract the dispersion relation from the peak of the spectral
	function $A(\boldsymbol{k},E)$ for a given $ \boldsymbol{k} $. The peak of $A(k=0.333/a,0.667/a,E)$
	moves toward $E=0$ as the disorder strength $\alpha$ increases, indicating that
	Dirac electron group velocity $v_{g}$ and the dispersion relationship are strongly
	renormalized due to the multi-scattering events. This is quite different from the
	usual picture in conventional metal with a finite density of states (DoS), where
	the life-time effects dominate.
	
	%This behavior indicates that the anomalous dimension of the Dirac field vanishes. This result is consistent with one-loop renormalization group calculation, note that our simulation is exact.
	
	%Fourth, from the calculated self-energy, the density of states $$ .

	\section{Density of states}
	\label{subsec.DoS}
	
	To make our results even more convincing, we revisit the calculation of the DoS based
	on our spectral function function. The single-particle DoS can be easily obtained through
	our simulated self-energy function as

	\begin{equation}
		\begin{aligned}
			\rho(E)= & -\frac{A_{c}}{2\pi} \int\frac{d^{2}\boldsymbol{k}}{(2\pi)^{2}} [\mathrm{Im}G^{R}(\boldsymbol{k}+,E) +\mathrm{Im}G^{R}(\boldsymbol{k}-,E)]
			\\= & \frac{A_{c}\eta}{2(\hbar v_{f})^{2}\pi^{2}} \left[\mathrm{ln}\frac{E_{c}^{2}}{a^{2}+\eta^2}
			+\frac{2a}{\eta}\mathrm{tan}^{-1}\left( \frac{a}{\eta}\right) \right].
		\end{aligned}
		\label{eq.dos}
	\end{equation}
	Here we directly compare this result with the average DoS obtained by the widely
	used Lanczos method in real space that has been well studied by our former work
	\cite{Wu}. As shown in Fig.~\ref{fig:self}, for a substantial energy range $[-0.1,0.1]$, the
	DoS calculated by our simulated self-energy agrees well with the results obtained
	by the real-space method. The line shape of DoS deviates from linearity to sub-linearity
	as the disorder strength increases, quite similar to the behavior of the imaginary part
	of the self-energy $\mathrm{Im}\Sigma(E)$. Furthermore, according to Eq.~(\ref{eq.dos})
	and our simulated results for $\Sigma_{0}$ in the main text, the DoS at the Dirac point
	$\rho(0)=A_{c}/(2\hbar^2 v_{f}^{2}\pi^{2})\cdot \Sigma_{0}/\alpha\sim \exp(-1/2\alpha)/\alpha$,
	which is also consistent with the results obtained by the functional Renormaliztion group
	technique \cite{Katanin,Sbierski}.

\section{Vertex Correction For Conductivity}
\label{subsec.vertex}

As mentioned in the main text, the vertex correction only contributes in the long range disorder
case. By only considering the intravalley scattering, the current vertex $ J_x $ satisfies the Beta-Salpeter equation of Eq.~(\ref{vertex-1}). In the vicinity of the single-valley Dirac point, we can also neglect the momentum-dependence
of the disorder potential correlator, and obtain
\begin{equation}
	\begin{aligned}
		\langle V_{\boldsymbol{k}-\boldsymbol{k}^{'}}V_{\boldsymbol{k}^{'}-\boldsymbol{k}}\rangle_{\text{dis}} \approx \gamma (\hbar v_{f})^{2}.
	\end{aligned}
	\label{current}
\end{equation}
Here, we adopt the Gaussian correlated disordered potential that has been described in the main text. The summation of the discrete momentum $\boldsymbol{k}$ in Eq. (\ref{vertex-1}) will be replaced by the integral of the first Brillion zone, i.e., $1/N\sum_{\boldsymbol{k}}\rightarrow A_{c}/(2\pi)^{2}\int d^{2}\boldsymbol{k}$, and then
\begin{equation}
	\begin{aligned}
		&\Lambda^{ss^{'}}(E)\sigma_{x}=  \sigma_{x}+\Lambda^{ss^{'}}(E)\gamma (\hbar v_{f})^{2} \int\frac{d^{2}\boldsymbol{k}^{'}}{(2\pi)^{2}} 
		\\& \times
		\frac{1}{a+is\eta-\hbar v_{f}\boldsymbol{k}^{'}\cdot \boldsymbol{\sigma}} \sigma_{x}\frac{1}{a+is\eta-\hbar v_{f}\boldsymbol{k}^{'}\cdot \boldsymbol{\sigma}}.
	\end{aligned}
	\label{betasalpter}
\end{equation}
With some algebraic operation, one can directly obtain
\begin{equation}
	\begin{aligned}\Lambda^{ss^{'}}(E)=[1-\frac{\gamma}{4\pi}\mathcal{I}(E,s,s^{'})]^{-1},\end{aligned}
	\label{prefactor}
\end{equation}
with
\begin{align}
	\begin{split}
	\mathcal{I}(E,s,s^{'})=&\int_{0}^{k_{c}}dk\ k \frac{2(\hbar v_{f})^{2} (a+is\eta)} {(a+is\eta)^{2}-(\hbar v_fk)^2}
	\\& \times \frac{(a+is^{'}\eta)} {(a+is^{'}\eta)^{2}-(\hbar v_fk)^2}.
	\end{split}
\end{align}
After performing the integration, we have
\begin{equation}
	\begin{aligned} & \mathcal{I}(E,+,+)=\mathcal{I}(E,-,-)=-1,\\
		& \mathcal{I}(E,+,-)=\mathcal{I}(E,-,+)=(\frac{a}{\eta}+\frac{\eta}{a})\mathrm{arctan}\frac{a}{\eta}.
	\end{aligned}
	\label{integral}
\end{equation}
Finally, we find that with the vertex correction, the Kubo formula for conductivity is
given by
\begin{equation}
	\begin{aligned}\sigma_{xx}(E)= & -\frac{\hbar e^{2}v_{f}^{2}}{4\pi}\sum_{ss^{'}=\pm}ss^{'}\int\frac{d^{2}\boldsymbol{k}}{(2\pi)^{2}}\mathrm{Tr}[\sigma_{x}G(\boldsymbol{k},a+is\eta)\\
		& \times J_{x}(\boldsymbol{k},a+is\eta,a+is^{'}\eta)G(\boldsymbol{k},a+is^{'}\eta)]\\
		= & \frac{e^{2}}{2\pi h}\frac{1+\mathcal{I}(E)}{(1+\gamma/4\pi)(1-\gamma/4\pi\mathcal{I}(E))},
	\end{aligned}
	\label{kubo}
\end{equation}
where $\mathcal{I}(E)=(a/\eta+\eta/a)\mathrm{arctan}(a/\eta)$. %Fig. \ref{fig:vertex} shows the examples of calculated $\sigma_{xx}(E)$ as a function of Fermi energy
%$E$ and for $\gamma=0.07,0.13$. We plot both the results with and without vertex correction as a comparison. 
At $E=0$, we have $\mathcal{I}(E=0)=1$, and then get
\begin{equation}
	\begin{aligned}\sigma_{xx}(E=0)=\frac{e^{2}}{\pi h}\frac{1}{1-(\frac{\gamma}{4\pi})^{2}}.\end{aligned}
	\label{mini}
\end{equation}
According to Eq.~(\ref{mini}), the minimum conductivity will be dependent on the disorder strength with the vertex correction. However, this dependence is extremely small. As one moves away from the Dirac
point, the vertex correction becomes large and gradually gets close to the result $\Lambda=2$
in the conventional metal.

\section{Quantum Interference Corrections For Conductivity}
\label{subsec.QI}
In this section, we calculate the quantum correction to the classical conductivity.
The low energy electron excitation of graphene is well described by the two-valley
massless Dirac model in two dimension that is given by
\begin{equation}
	H_{s}=\hbar v_f (sk_{x}\sigma_{x}-k_{y}\sigma_{y}),
\end{equation}
where $s=\pm$ stands for $K$ and $K^{\prime}$ valleys, respectively. We suppose that
the Fermi level $E_{f}$ intersects the conduction band with the dispersion as
\begin{equation}
	\epsilon_{\boldsymbol{k}}=\hbar v_f |\boldsymbol{k}|,
\end{equation}
and the corresponding eigenfunctions are
\begin{equation}
	\langle\boldsymbol{r}|s\boldsymbol{k}\rangle=\psi_{s\boldsymbol{k}}(\boldsymbol{r})=\frac{1}{\sqrt{2S}}\left[\begin{array}{c}
		1\\
		se^{-is\theta_{\boldsymbol{k}}}
	\end{array}\right]e^{i\boldsymbol{k} \cdot\boldsymbol{r}}.
\end{equation}
The disorder-induced self-energy is obtained numerically through the momentum-space Lanczos
methods introduced in the main text, and then the retarded (R) and advanced (A) Green's
functions have the form
\begin{equation}
	G_{K,K^{\prime}}^{R/A}(\boldsymbol{k},\omega)=\frac{1}{\omega-\epsilon_{\boldsymbol{k}}-\mathrm{Re}\Sigma^{R}(\omega)\mp i\mathrm{Im}\Sigma^{R}(\omega)}.
\end{equation}

In order to evaluate the quantum correction to the classical conductivity, we need
to calculate a summation of maximally crossed diagrams, which is denoted by
\begin{equation}
	\sigma_{qi}=\sigma_{KK}^{KK}+\sigma_{K^{\prime}K^{\prime}}^{K^{\prime}K^{\prime}}+\sigma_{K^{\prime}K}^{KK^{\prime}}+\sigma_{KK^{\prime}}^{K^{\prime}K},\label{eq:totalsigma}
\end{equation}
with %
\begin{widetext}
	\begin{align}
		\sigma_{KK}^{KK} & =\frac{e^{2}\hbar}{2\pi S}\sum_{\boldsymbol{k}}\sum_{\boldsymbol{q}}\varGamma_{KK}^{KK}(\theta_{\boldsymbol{k}},\theta_{-\boldsymbol{k}},\boldsymbol{q})G_{K}^{R}(\boldsymbol{k})v_{K}^{x}(\boldsymbol{k})G_{K}^{A}(\boldsymbol{k})G_{K}^{R}(\boldsymbol{q}-\boldsymbol{k})v_{K}^{x}(\boldsymbol{q}-\boldsymbol{k})G_{K}^{A}(\boldsymbol{q}-\boldsymbol{k}),\label{eq:sigma1}\\
		\sigma_{\bar{K}K}^{K\bar{K}} & =\frac{e^{2}\hbar}{2\pi S}\sum_{\boldsymbol{k}}\sum_{\boldsymbol{q}}\varGamma_{\bar{K}K}^{K\bar{K}}(\boldsymbol{q})G_{K}^{R}(\boldsymbol{k})v_{K}^{x}(\boldsymbol{k})G_{K}^{A}(\boldsymbol{k})G_{\bar{K}}^{R}(\boldsymbol{q}-\boldsymbol{k})v_{\bar{K}}^{x}(\boldsymbol{q}-\boldsymbol{k})G_{\bar{K}}^{A}(\boldsymbol{q}-\boldsymbol{k}).\label{eq:sigma2}
	\end{align}
\end{widetext}

There exists three types of Cooperon (particle-particle type) channels and the full
vertex function $\varGamma$ is related to $\gamma$ by the Bethe-Salpter equation:
\begin{widetext}
	\begin{align}
		\varGamma_{KK}^{KK}(\theta_{\boldsymbol{p}},\theta_{\boldsymbol{p}^{\prime}};\boldsymbol{q})&=\gamma_{KK}^{KK}(\theta_{\boldsymbol{p}},\theta_{\boldsymbol{p}^{\prime}})+\frac{1}{S}\sum_{\boldsymbol{k}}\gamma_{KK}^{KK}(\theta_{\boldsymbol{p}},\theta_{\boldsymbol{k}})G_{K}^{R}(\boldsymbol{k})G_{K}^{A}(\boldsymbol{q}-\boldsymbol{k})\varGamma_{KK}^{KK}(\theta_{\boldsymbol{k}},\theta_{\boldsymbol{p}^{\prime}};\boldsymbol{q}),\label{eq:BSE_for_intra_Cooperon_channel}
		\\
		\varGamma_{\bar{K}\bar{K}}^{KK}(\theta_{\boldsymbol{p}},\theta_{\boldsymbol{p}^{\prime}};\boldsymbol{q}) & =\gamma_{\bar{K}\bar{K}}^{KK}(\theta_{\boldsymbol{p}},\theta_{\boldsymbol{p}^{\prime}})+\frac{1}{S}\sum_{\boldsymbol{k}}\Big[\gamma_{\bar{K}\bar{K}}^{KK}(\theta_{\boldsymbol{p}},\theta_{\boldsymbol{k}})G_{K}^{R}(\boldsymbol{k})G_{\bar{K}}^{A}(\boldsymbol{q}-\boldsymbol{k})\varGamma_{\bar{K}\bar{K}}^{KK}(\theta_{\boldsymbol{k}},\theta_{\boldsymbol{p}^{\prime}};\boldsymbol{q})\nonumber \\
		&\quad +\gamma_{\bar{K}K}^{K\bar{K}}(\theta_{\boldsymbol{p}},\theta_{\boldsymbol{k}})G_{\bar{K}}^{R}(\boldsymbol{k})G_{K}^{A}(\boldsymbol{q}-\boldsymbol{k})\varGamma_{K\bar{K}}^{\bar{K}K}(\theta_{\boldsymbol{k}},\theta_{\boldsymbol{p}^{\prime}};\boldsymbol{q})\Big]\label{eq:inter_channel_1},\\
		\varGamma_{\bar{K}K}^{K\bar{K}}(\theta_{\boldsymbol{p}},\theta_{\boldsymbol{p}^{\prime}};\boldsymbol{q}) & =\gamma_{\bar{K}K}^{K\bar{K}}(\theta_{\boldsymbol{p}},\theta_{\boldsymbol{p}^{\prime}})+\frac{1}{S}\sum_{\boldsymbol{k}}\Big[\gamma_{\bar{K}\bar{K}}^{KK}(\theta_{\boldsymbol{p}},\theta_{\boldsymbol{k}})G_{K}^{R}(\boldsymbol{k})G_{\bar{K}}^{A}(\boldsymbol{q}-\boldsymbol{k})\varGamma_{\bar{K}K}^{K\bar{K}}(\theta_{\boldsymbol{k}},\theta_{\boldsymbol{p}^{\prime}};\boldsymbol{q})\nonumber \\
		&\quad +\gamma_{\bar{K}K}^{K\bar{K}}(\theta_{\boldsymbol{p}},\theta_{\boldsymbol{k}})G_{\bar{K}}^{R}(\boldsymbol{k})G_{K}^{A}(\boldsymbol{q}-\boldsymbol{k})\varGamma_{KK}^{\bar{K}\bar{K}}(\theta_{\boldsymbol{k}},\theta_{\boldsymbol{p}^{\prime}};\boldsymbol{q})\Big]\label{eq:inter_channel_2}.
	\end{align}
\end{widetext}
where $\theta_{\boldsymbol{p}}$ and $\theta_{\boldsymbol{p}^{\prime}}$ label the incoming
and outgoing momenta, respectively, and we have neglected the $\boldsymbol{q}$ dependence in the bare
scattering vertex. The bare scattering vertex which only causes small momentum transfer
within the single valley can be expressed as

\begin{align}
	\begin{split}
		&\gamma_{KK}^{KK}(\theta_{\boldsymbol{p}},\theta_{\boldsymbol{p}^{\prime}}) 
		\\& =\frac{(\hbar v_f)^{2}}{E\tau_{0}/\hbar}\langle K,\boldsymbol{p}|K,\boldsymbol{p}^{\prime}\rangle\langle K,-\boldsymbol{p}|K,-\boldsymbol{p}^{\prime}\rangle
		\\&=\frac{(\hbar v_f)^{2}}{2E\tau_{0}/\hbar}[\frac{1}{2}e^{-2i(\theta-\theta^{\prime})}+e^{-i(\theta-\theta^{\prime})}+\frac{1}{2}],
	\end{split}
\end{align}
and
\begin{align}
	\begin{split}
		&\gamma_{\bar{K}\bar{K}}^{KK}  =\gamma_{KK}^{\bar{K}\bar{K}}
		\\&=\frac{(\hbar v_f)^{2}}{E\tau_{0}/\hbar}\langle K,\boldsymbol{p}|K,\boldsymbol{p}^{\prime}\rangle\langle K^{\prime},-\boldsymbol{p}|K^{\prime},-\boldsymbol{p}^{\prime}\rangle
		\\&=\frac{(\hbar v_f)^{2}}{2E\tau_{0}/\hbar}[1+\frac{1}{2}e^{i(\theta-\theta^{\prime})}+\frac{1}{2}e^{-i(\theta-\theta^{\prime})}].
	\end{split}
\end{align}
The bare scattering vertex which causes the scattering of electrons between two valleys
can be expressed as
\begin{align}
	\begin{split}
		&\gamma_{K^{\prime}K}^{KK^{\prime}}=\gamma_{KK^{\prime}}^{K^{\prime}K} 
		\\& =\frac{(\hbar v_f)^{2}}{E\tau_{i}/\hbar}\langle K,\boldsymbol{p}|K^{\prime},\boldsymbol{p}^{\prime}\rangle\langle K^{\prime},-\boldsymbol{p}|K-\boldsymbol{p}^{\prime}\rangle
		\\&=\frac{(\hbar v_f)^{2}}{2E\tau_{i}/\hbar}[1-\frac{1}{2}e^{i(\theta+\theta^{\prime})}-\frac{1}{2}e^{-i(\theta+\theta^{\prime})}].
	\end{split}
\end{align}
We have introduced intra- and inter-disorder strengths of $\frac{(\hbar v_f)^{2}}{E\tau_{0}/\hbar}$ and $\frac{(\hbar v_f)^{2}}{E\tau_{i}/\hbar}$. The total disorder strength is given in terms of $\tau_{0}$ and $\tau_{i}$,
\begin{align}
	\frac{(\hbar v_f)^{2}}{E\tau_{t}/\hbar}=\frac{(\hbar v_f)^{2}}{E\tau_{0}/\hbar}+\frac{(\hbar v_f)^{2}}{E\tau_{i}/\hbar}.
\end{align}
As shown in Eqs.~(\ref{eq:BSE_for_intra_Cooperon_channel})-(\ref{eq:inter_channel_2}), the radial coordinate $k$ is only contained in the kernel $\frac{1}{S}\sum_{\boldsymbol{k}}G_{K}^{R}(\boldsymbol{k})G_{\bar{K}}^{A}(\boldsymbol{q}-\boldsymbol{k})$
and can be evaluated as
\begin{align}
	\begin{split}
		&\int_{0}^{\infty}\frac{dkk}{2\pi}G^{R}(\boldsymbol{k}+\frac{\boldsymbol{q}}{2})G^{A}(\frac{\boldsymbol{q}}{2}-\boldsymbol{k})
		\\\approx&\frac{\Pi(E)}{(\hbar v_f)^{2}}\left\{ 1-i(v_{g}\tau q)\cos\theta-(v_{g}\tau q)^{2}\cos^{2}\theta\right\}, \label{eq:radial_integration}
	\end{split}
\end{align}
with $ \Pi(E) =\frac{\chi(E)}{2\pi} \left(\frac{\pi}{2}+ \arctan\chi(E)\right) $.

For later convenience, we introduce the renormalized relaxation time through $\Pi(E)\equiv E\tau^{*}/\hbar$.
The angular coordinate can be done by using the expansion of the full vertex function $\varGamma$ and the bare vertex $\gamma$:
\begin{align}
	&\varGamma(\theta_{\boldsymbol{p}},\theta_{\boldsymbol{p}^{\prime}};\boldsymbol{q})  =\frac{(\hbar v_f)^{2}}{E\tau_{t}/\hbar}\sum_{n,m}\varGamma_{nm}(\boldsymbol{q})e^{i(n\theta_{\boldsymbol{p}}-m\theta_{\boldsymbol{p}^{\prime}})},\label{eq:full_vertex}\\&
	\gamma(\theta_{\boldsymbol{p}},\theta_{\boldsymbol{p}^{\prime}})  =\frac{(\hbar v_f)^{2}}{E\tau_{t}/\hbar}\sum_{n,m}\gamma_{nm}e^{i(n\theta_{\boldsymbol{p}}-m\theta_{\boldsymbol{p}^{\prime}})}.
\end{align}
If we further define
\begin{align}
	\begin{split}
		\Phi_{nm}&=\frac{1}{2\pi}\int_{0}^{2\pi}d\theta\ e^{i(m-n)\theta}
		\\& \quad \times\left\{ 1-i(v_{g}\tau q)\cos\theta-(v_{g}\tau q)^{2}\cos^{2}\theta\right\} ,
	\end{split}
\end{align}
the expansion coefficients in Eqs. (\ref{eq:BSE_for_intra_Cooperon_channel})-(\ref{eq:inter_channel_2}) can be expressed in the matrix form
\begin{align}
	\begin{split}
		\boldsymbol{\varGamma}_{KK}^{KK} & =\boldsymbol{\gamma}_{KK}^{KK}+\boldsymbol{\gamma}_{KK}^{KK}\boldsymbol{\Phi}\boldsymbol{\varGamma}_{KK}^{KK}, \\
		\boldsymbol{\varGamma}_{\bar{K}\bar{K}}^{KK} & =\boldsymbol{\gamma}_{\bar{K}\bar{K}}^{KK}+\boldsymbol{\gamma}_{\bar{K}\bar{K}}^{KK}\boldsymbol{\Phi}\boldsymbol{\varGamma}_{\bar{K}\bar{K}}^{KK}+\boldsymbol{\gamma}_{\bar{K}K}^{K\bar{K}}\boldsymbol{\Phi}\boldsymbol{\varGamma}_{K\bar{K}}^{\bar{K}K}, \\
		\boldsymbol{\varGamma}_{\bar{K}K}^{K\bar{K}} & =\boldsymbol{\gamma}_{\bar{K}K}^{K\bar{K}}+\boldsymbol{\gamma}_{\bar{K}\bar{K}}^{KK}\boldsymbol{\Phi}\boldsymbol{\varGamma}_{\bar{K}K}^{K\bar{K}}+\boldsymbol{\gamma}_{\bar{K}K}^{K\bar{K}}\boldsymbol{\Phi}\boldsymbol{\varGamma}_{KK}^{\bar{K}\bar{K}},\label{eq:matrix_form_BSE}
	\end{split}
\end{align}
where the bare scattering vertices are
\begin{align}
	\begin{split}
		&\boldsymbol{\gamma}_{KK}^{KK}=\frac{\tau}{\tau_{0}}\left[\begin{array}{ccc}
			\frac{1}{2} & 0 & 0\\
			0 & 1 & 0\\
			0 & 0 & \frac{1}{2}
		\end{array}\right],
		\\&\boldsymbol{\gamma}_{\bar{K}K}^{K\bar{K}}=\frac{\tau}{\tau_{i}}\left[\begin{array}{ccc}
			0 & 0 & -\frac{1}{2}\\
			0 & 1 & 0\\
			-\frac{1}{2} & 0 & 0
		\end{array}\right],
		\\&\boldsymbol{\gamma}_{\bar{K}\bar{K}}^{KK}=\frac{\tau}{\tau_{0}}\left[\begin{array}{ccc}
			\frac{1}{2} & 0 & 0\\
			0 & 1 & 0\\
			0 & 0 & \frac{1}{2}
		\end{array}\right].
	\end{split}
\end{align}
By truncating up to $q^{2}$ terms in small $q$ limit, $\boldsymbol{\Phi}$ has the form,
\begin{equation}
	\boldsymbol{\Phi}=\frac{\tau^{*}}{\tau_{t}}\left[\begin{array}{ccc}
		1-\frac{1}{2}(v_{g}\tau)^{2}q^{2} & -\frac{1}{2}iv_{g}\tau q_{+} & -\frac{1}{4}\ell_{e}^{2}q_{+}^{2}\\
		-\frac{1}{2}iv_{g}\tau q_{-} & 1-\frac{1}{2}(v_{g}\tau)^{2}q^{2} & -\frac{1}{2}iv_{g}\tau q_{+}\\
		-\frac{1}{4}\ell_{e}^{2}q_{-}^{2} & -\frac{1}{2}iv_{g}\tau q_{-} & 1-\frac{1}{2}(v_{g}\tau)^{2}q^{2}
	\end{array}\right],
\end{equation}
with $q^{2}=q_{x}^{2}+q_{y}^{2}$ and $q_{\pm}=q_{x}\pm iq_{y}$.%
The two Cooperon channels $\boldsymbol{\varGamma}_{\bar{K}\bar{K}}^{KK}$ and $\boldsymbol{\varGamma}_{\bar{K}K}^{K\bar{K}}$ in Eq. (\ref{eq:matrix_form_BSE}) are coupled together. By introducing the new variables,
\begin{align}
	\begin{split}
		\boldsymbol{x} & =\boldsymbol{\gamma}_{\bar{K}\bar{K}}^{KK}+\boldsymbol{\gamma}_{\bar{K}K}^{K\bar{K}},\;\;
		\\ \boldsymbol{y}&=\boldsymbol{\gamma}_{\bar{K}\bar{K}}^{KK}-\boldsymbol{\gamma}_{\bar{K}K}^{K\bar{K}},\;\;
		\\
		\boldsymbol{z}&=\boldsymbol{\gamma}_{KK}^{KK},
	\end{split}
	\\
	\begin{split}
		\boldsymbol{X} & =\boldsymbol{\varGamma}_{\bar{K}\bar{K}}^{KK}+\boldsymbol{\varGamma}_{\bar{K}K}^{K\bar{K}},\;\;
		\\
		\boldsymbol{Y}&=\boldsymbol{\varGamma}_{\bar{K}\bar{K}}^{KK}-\boldsymbol{\varGamma}_{\bar{K}K}^{K\bar{K}},\;\;
		\\
		\boldsymbol{Z}&=\boldsymbol{\varGamma}_{KK}^{KK}\label{eq:inter_cooperon_X},
	\end{split}
\end{align}
the coupled Bethe-Salpeter equations (Eq.~(\ref{eq:matrix_form_BSE})) are reduced to uncoupled
ones and the expansion coefficients can be easily solved through
\begin{align}
	\begin{split}
		&\boldsymbol{X}  =\left[\boldsymbol{1}_{3\times3}-\boldsymbol{x}\boldsymbol{\Phi}\right]^{-1}\boldsymbol{x},\;\;
		\\&\boldsymbol{Y}=\left[\boldsymbol{1}_{3\times3}-\boldsymbol{y}\boldsymbol{\Phi}\right]^{-1}\boldsymbol{y},\;\;
		\\&\boldsymbol{Z}=\left[\boldsymbol{1}_{3\times3}-\boldsymbol{z}\boldsymbol{\Phi}\right]^{-1}\boldsymbol{z}.
	\end{split}
\end{align}
By retaining the most singular terms, we can solve the above three matrix equations:
\begin{align}
	\begin{split}
		&\boldsymbol{X}  \approx\left[\begin{array}{ccc}
			0 & 0 & 0\\
			0 & \frac{1}{g_{x}+D_{ter}\tau q^{2}} & 0\\
			0 & 0 & 0
		\end{array}\right],\;\;
		\\&\boldsymbol{Y}\approx\left[\begin{array}{ccc}
			0 & 0 & 0\\
			0 & \frac{1}{g_{y}+D_{ter}\tau q^{2}} & 0\\
			0 & 0 & 0
		\end{array}\right],\;\;
		\\&\boldsymbol{Z}\approx\left[\begin{array}{ccc}
			0 & 0 & 0\\
			0\frac{1}{2} & \frac{1}{g_{z}+D_{tra}\tau q^{2}} & 0\\
			0 & 0 & 0
		\end{array}\right],
	\end{split}
\end{align}
with the Cooperon gaps
\begin{align}
	\begin{split} 
		&g_{x}=1-\frac{\tau^{*}}{\tau_{t}},
		\\& g_{y}=(\frac{\tau_{t}}{\tau_{0}}-\frac{\tau_{t}}{\tau_{i}})^{-1}-\frac{\tau^{*}}{\tau_{t}},
		\\& g_{z}=\frac{\tau_{0}-\tau^{*}}{2\tau_{t}},
		\label{eq:cooperon_gaps}
	\end{split}
\end{align}
and the diffusive constants for the inter- and intra-Cooperon channels
\begin{align}
	\begin{split}
	D_{ter} & =\left[\left(2\frac{\tau}{\tau^{*}}-\frac{\tau_{t}}{\tau_{0}}+\frac{\tau_{t}}{\tau_{i}}\right)^{-1}+\left(2\frac{\tau_{t}}{\tau^{*}}-1\right)^{-1}\right]\frac{v_{g}^{2}\tau}{2},\\
	D_{tra} & =\frac{\tau^{*}}{\tau_{t}}(1+\frac{\tau_{t}}{\tau_{i}})^{-1}\frac{v_{g}^{2}\tau}{2}.
	\end{split}
\end{align}
Thus, according to Eq. (\ref{eq:inter_cooperon_X}), these Cooperons are evaluated as
\begin{align}
	\varGamma_{\bar{K}K}^{K\bar{K}}(\theta_{\boldsymbol{p}},\theta_{\boldsymbol{p}^{\prime}};\boldsymbol{q}) & =\frac{1}{2}\frac{(\hbar v_f)^{2}}{E\tau_{t}/\hbar}\left(\frac{1}{g_{x}+D_{ter}\tau q^{2}}-\frac{1}{g_{y}+D_{ter}\tau q^{2}}\right),\label{eq:Cooperon_propagator}\\
	\varGamma_{\bar{K}\bar{K}}^{KK}(\theta_{\boldsymbol{p}},\theta_{\boldsymbol{p}^{\prime}};\boldsymbol{q}) & =\frac{1}{2}\frac{(\hbar v_f)^{2}}{E\tau_{t}/\hbar}\left(\frac{1}{g_{x}+D_{ter}\tau q^{2}}+\frac{1}{g_{y}+D_{ter}\tau q^{2}}\right),\\
	\varGamma_{KK}^{KK}(\theta_{\boldsymbol{p}},\theta_{\boldsymbol{p}^{\prime}};\boldsymbol{q}) & =\frac{1}{2}\frac{(\hbar v_f)^{2}}{E\tau_{t}/\hbar}\frac{e^{i(\theta_{\boldsymbol{p}}-\theta_{\boldsymbol{p}^{\prime}})}}{g_{z}+D_{tra}\tau q^{2}}.
\end{align}
Generally speaking, the total quantum correction is determined by all these four Cooperon channels (the intra-valley Cooperon channels are doubly degenerate). But,
we are interested in how the multi-scattering effects will qualitatively renormalize the quantum interference correction to the conductivity. In the following, we only
discuss two limiting regimes with two different types of scattering.

\subsection{Short range disorder}
For short range impurities, $2\tau_{t}=\tau_{0}=\tau_{i}$, and the intra-valley Cooperon channel $\boldsymbol{\varGamma}_{KK}^{KK}$ is always fully gapped with its contribution suppressed. In this situations, we only need to consider the contribution from the inter-valley Cooperon channel $\boldsymbol{\varGamma}_{\bar{K}K}^{K\bar{K}}$.
From Eq. (\ref{eq:sigma2}), we first evaluate the bare Hikami box. Since here we consider that the external momentum is zero, the bare Hikami box for the Cooperon
channel $\boldsymbol{\varGamma}_{\bar{K}\bar{K}}^{KK}$ vanishes, and the bare Hikami box contribution for the inter-valley Cooperon channel is %
\begin{widetext}
	\begin{align}
		\begin{split}
			\sigma_{ter}^{qi(0)}= & \frac{e^{2}\hbar}{2\pi}\int\frac{d^{2}\boldsymbol{q}}{(2\pi)^{2}}\int\frac{d^{2}\boldsymbol{k}}{(2\pi)^{2}}G_{\boldsymbol{K}}^{R}(\boldsymbol{k})v_{\boldsymbol{K}}^{x}(\boldsymbol{k})G_{\boldsymbol{K}}^{A}(\boldsymbol{k})G_{\boldsymbol{K}}^{R}(-\boldsymbol{k})v_{\boldsymbol{K}}^{x}(-\boldsymbol{k})G_{\boldsymbol{K}}^{A}(-\boldsymbol{k})\varGamma_{\bar{K}K}^{K\bar{K}}(\theta_{\boldsymbol{k}},\theta_{-\boldsymbol{k}};\boldsymbol{q})\\
			= & -\frac{e^{2}}{2\pi h}\frac{\frac{1}{2\pi}+\Pi(E)}{E\tau_{t}/\hbar}\ln\frac{D_{ter}\tau/\ell_{e}^{2}+g_{x}}{D_{ter}\tau/\ell_{\phi}^{2}+g_{x}}.
		\end{split}
	\end{align}
\end{widetext}
The full correction to the conductivity should take into account the dressed Hikami box contribution. It is reported to have the same order of magnitude of the bare Hikami box and different signs in two-dimensional systems with large spin-orbital coupling \cite{McCann}.

For the inter-valley Cooperon channels, we need to consider the following dressed Hikami box contribution,
\begin{widetext}
	\begin{align}
		\begin{split}
			\sigma_{ter}^{qi(1)} & =2\frac{e^{2}\hbar}{2\pi}\int\frac{d^{2}\boldsymbol{q}}{(2\pi)^{2}}\int\frac{d^{2}\boldsymbol{k}}{(2\pi)^{2}}\int\frac{d^{2}\boldsymbol{p}}{(2\pi)^{2}}\varGamma_{\bar{K}\bar{K}}^{KK}(\theta_{\boldsymbol{k}},\theta_{\boldsymbol{p}};\boldsymbol{q})\langle U_{-\boldsymbol{k},\boldsymbol{p}}^{\boldsymbol{K}\bar{\boldsymbol{K}}}U_{\boldsymbol{k},-\boldsymbol{p}}^{\boldsymbol{\bar{K}K}}\rangle_{\text{imp}}\\
			& \quad \times G_{\bar{\boldsymbol{K}}}^{R}(-\boldsymbol{k})G_{\bar{\boldsymbol{K}}}^{R}(\boldsymbol{p})v_{\bar{\boldsymbol{K}}}^{x}(\boldsymbol{p})G_{\bar{\boldsymbol{K}}}^{A}(\boldsymbol{p})G_{\bar{\boldsymbol{K}}}^{R}(-\boldsymbol{p})G_{\bar{\boldsymbol{K}}}^{R}(\boldsymbol{k})v_{\bar{\boldsymbol{K}}}^{x}(\boldsymbol{k})G_{\bar{\boldsymbol{K}}}^{A}(\boldsymbol{k})\\
			& =-\frac{e^{2}}{2\pi h}\frac{(\frac{1}{2\pi}+\Pi(E))^{2}}{4(E\tau_{t}/\hbar) (E\tau_{i}/\hbar)} \ln\frac{D_{ter}\tau/\ell_{e}^{2}+g_{x}}{D_{ter}\tau/\ell_{\phi}^{2}+g_{x}},
		\end{split}
%	\end{align}
%	\begin{align}
		\\
		\begin{split}
			\sigma_{ter}^{qi(2)} & =2\frac{e^{2}\hbar}{2\pi}\int\frac{d^{2}\boldsymbol{q}}{(2\pi)^{2}}\int\frac{d^{2}\boldsymbol{k}}{(2\pi)^{2}}\int\frac{d^{2}\boldsymbol{p}}{(2\pi)^{2}}\varGamma_{\bar{K}K}^{K\bar{K}}(\theta_{\boldsymbol{k}},\theta_{\boldsymbol{p}};\boldsymbol{q})\langle U_{-\boldsymbol{k},\boldsymbol{p}}^{\boldsymbol{K}\boldsymbol{K}}U_{\boldsymbol{k},-\boldsymbol{p}}^{\boldsymbol{KK}}\rangle_{\text{imp}}\\
			& \quad \times G_{\boldsymbol{K}}^{R}(-\boldsymbol{k})G_{\boldsymbol{K}}^{R}(\boldsymbol{p})v_{\boldsymbol{K}}^{x}(\boldsymbol{p})G_{\boldsymbol{K}}^{A}(\boldsymbol{p})G_{\bar{\boldsymbol{K}}}^{R}(-\boldsymbol{p})G_{\bar{\boldsymbol{K}}}^{R}(\boldsymbol{k})v_{\bar{\boldsymbol{K}}}^{x}(\boldsymbol{k})G_{\bar{\boldsymbol{K}}}^{A}(\boldsymbol{k})\\
			& =\frac{e^{2}}{2\pi h}\frac{(\frac{1}{2\pi}+\Pi(\mu))^{2}}{4(E\tau_{t}/\hbar)(E\tau_{0}/\hbar)}\ln\frac{D_{ter}\tau/\ell_{e}^{2}+g_{x}}{D_{ter}\tau/\ell_{\phi}^{2}+g_{x}}.
		\end{split}
	\end{align}
\end{widetext}
After collecting all these contributions, we finally obtain the quantum interference correction for the inter-valley Cooperon channel as
\begin{align}
	\begin{split}
		\sigma_{ter}^{qi}&=\sigma_{ter}^{qi(0)}+\sigma_{ter}^{qi(1)}+\sigma_{ter}^{qi(2)}
		\\&=-\frac{e^{2}}{2\pi h}\frac{\frac{1}{2\pi}+\Pi(E)}{E\tau_{t}/\hbar}\ln\frac{D_{ter}\tau/\ell_{e}^{2}+g_{x}}{D_{ter}\tau/\ell_{\phi}^{2}+g_{x}}.
	\end{split}
\end{align}
If the chemical potential is located far from the Dirac node, we have $\Pi(E)\sim E\tau/\hbar\gg1$ and the Cooperon gap $g_{x}=1-\frac{\tau^{*}}{\tau_{t}}$ vanishes since $\tau^{*}\sim\tau_{t}$. The quantum interference conductivity correction is $\sigma_{ter}^{qi}=-\frac{e^{2}}{\pi h}\ln\frac{\ell_{\phi}}{\ell_{e}}$, recovering the results of the conventional weak localization regime. When the
chemical potential is near the Dirac point (strong scattering regime), due to the finite Cooperon gap ($g_{x}\approx1$), the quantum interference correction is strongly suppressed.

\subsection{Long range disorder}

For the long range potential disorder ($\tau_{t}\approx\tau_{0}\ll\tau_{i}$), the inter-valley Cooperon channel $\varGamma_{\bar{K}K}^{K\bar{K}}$ directly vanishes since $g_{x}\approx g_{y}$ and the channel $\varGamma_{\bar{K}\bar{K}}^{KK}$ can also be neglected since it is proportional to the inter-valley scattering strength.
Thus, only the intra-valley channel $\varGamma_{KK}^{KK}$ will contribute to the quantum interference correction. The bare Hikami box for the intra-valley Cooperon channel can be evaluated as
\begin{widetext}
	\begin{align}
		\begin{split}
			\sigma_{tra}^{qi(0)}= & 2\times\frac{e^{2}\hbar}{2\pi}\int\frac{d^{2}\boldsymbol{q}}{(2\pi)^{2}}\int\frac{d^{2}\boldsymbol{k}}{(2\pi)^{2}}G_{\boldsymbol{K}}^{R}(\boldsymbol{k})v_{\boldsymbol{K}}^{x}(\boldsymbol{k})G_{\boldsymbol{K}}^{A}(\boldsymbol{k})G_{\boldsymbol{K}}^{R}(-\boldsymbol{k})v_{\boldsymbol{K}}^{x}(-\boldsymbol{k})G_{\boldsymbol{K}}^{A}(-\boldsymbol{k})\varGamma_{KK}^{KK}(\theta_{\boldsymbol{k}},\theta_{-\boldsymbol{k}};\boldsymbol{q})\\
			\approx & \frac{e^{2}}{2\pi h}\frac{\frac{1}{2\pi}+\Pi(\mu)}{E\tau_{t}/\hbar } \ln\frac{D_{tra}\tau/\ell_{e}^{2}+g_{z}}{D_{tra}\tau/\ell_{\phi}^{2}+g_{z}},
		\end{split}
	\end{align}
\end{widetext}
where the prefactor 2 is due to the degeneracy of the intra-valley Cooperon channel.
The phase factor $e^{i(\theta_{\boldsymbol{k}}-\theta_{-\boldsymbol{k}})}=-1$ gives an additional minus sign compared with the inter-valley Cooperon channels (\ref{eq:Cooperon_propagator})
due to the $\pi$ berry phase. The dressed Hikami box contributions for the intra-valley Cooperon channels are
\begin{widetext}
	\begin{align}
		\begin{split}
		\sigma_{tra}^{qi(1)} & =4\frac{e^{2}}{2\pi h}\int\frac{d^{2}\boldsymbol{q}}{(2\pi)^{2}}\int\frac{d^{2}\boldsymbol{k}}{(2\pi)^{2}}\int\frac{d^{2}\boldsymbol{p}}{(2\pi)^{2}}\varGamma_{KK}^{KK}(\theta_{\boldsymbol{k}},\theta_{\boldsymbol{p}};\boldsymbol{q})\langle U_{-\boldsymbol{k},\boldsymbol{p}}^{\boldsymbol{K}\boldsymbol{K}}U_{\boldsymbol{k},-\boldsymbol{p}}^{\boldsymbol{KK}}\rangle_{imp}\\
		& \quad\times G_{\boldsymbol{K}}^{R}(-\boldsymbol{k})G_{\boldsymbol{K}}^{R}(\boldsymbol{p})v_{\boldsymbol{K}}^{x}(\boldsymbol{p})G_{\boldsymbol{K}}^{A}(\boldsymbol{p})G_{\boldsymbol{K}}^{R}(-\boldsymbol{p})G_{\boldsymbol{K}}^{R}(\boldsymbol{k})v_{\boldsymbol{K}}^{x}(\boldsymbol{k})G_{\boldsymbol{K}}^{A}(\boldsymbol{k})\\
		& =-\frac{e^{2}}{2\pi h}\frac{(\frac{1}{2\pi}+\Pi(E))^{2}}{2(E\tau_{t}/\hbar)^{2}}\ln\frac{D_{tra}\tau/\ell_{e}^{2}+g_{z}}{D_{tra}\tau/\ell_{\phi}^{2}+g_{z}}.
		\end{split}
	\end{align}
\end{widetext}
Then, we can obtain the quantum interference correction for the intra-valley Cooperon
channel as
\begin{align}
	\begin{split}
		\sigma_{tra}^{qi}&=\sigma_{tra}^{qi(0)}+\sigma_{tra}^{qi(1)}
		\\&=\frac{e^{2}}{2\pi h}\frac{\frac{1}{2\pi}+\Pi(E)}{E\tau_{t}/\hbar}\left[1-\frac{\frac{1}{2\pi}+\Pi(E)}{2E\tau_{t}/\hbar}\right]
		\\&\quad \times \ln\frac{D_{tra}\tau/\ell_{e}^{2}+g_{z}}{D_{tra}\tau/\ell_{\phi}^{2}+g_{z}}.
	\end{split}
\end{align}
In the weak scattering regime $(E\tau/\hbar\gg1)$, the Cooperon gap $g_{z}=\frac{1}{2}\left(1-\frac{\tau^{*}}{\tau_{t}}\right)$
vanishes since $\tau^{*}\sim\tau_{t}$ . In this situation, after including the vertex correcion, the quantum interference
conductivity correction is $\sigma_{ter}^{qi}=\frac{2e^{2}}{\pi h}\ln\frac{\ell_{\phi}}{\ell_{e}}$,
recovering the results of the weak anti-localization for the symplectic symmetry
class. When the chemical potential is near the Dirac point (strong scattering regime),
the Cooperon gap is finite ($g_{z}\approx\frac{1}{2}$), the quantum interference correction
will be strongly suppressed.

%\section{Self-energy With Correlated Gaussian Potential}
\section{Linear fittings of $\beta$ and $\Delta$ with the correlated Gaussian potential}
\label{subsec.Gaussian}

\break

\begin{figure}[t!]
	\includegraphics[width=8.5cm]{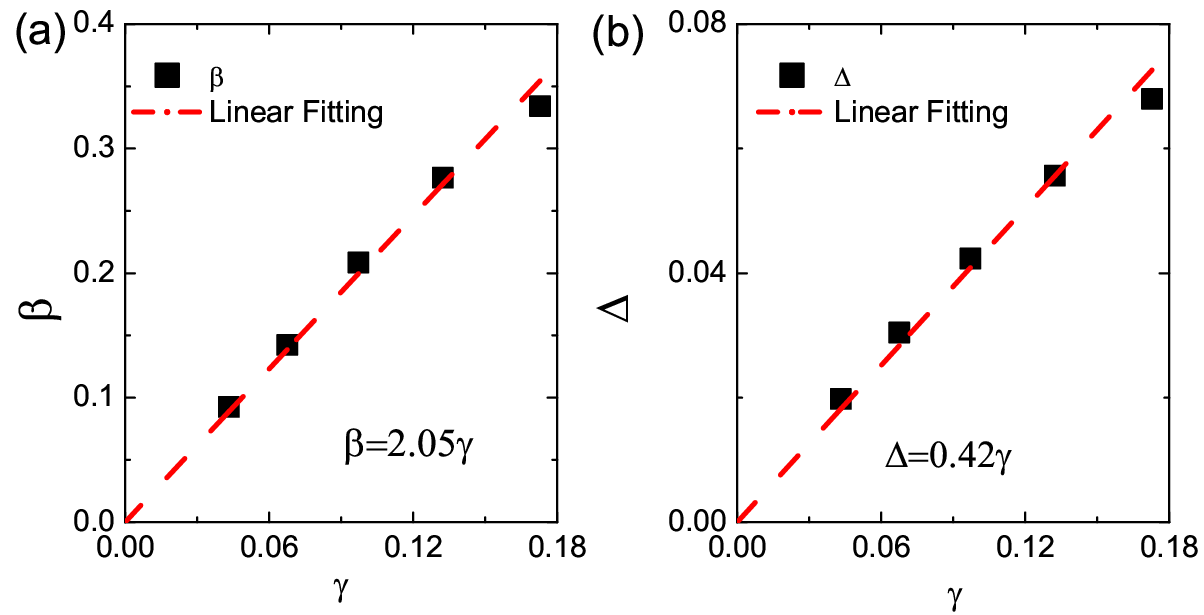} \caption{Linear fittings of (a) $\beta$ and (b) $\Delta$ as functions of the disorder strength ($\gamma$) for the correlation length $\xi=2.0a$. The black squares denote the data obtained from the power-law fitting self-energy.}
	\label{fig:parameter}
\end{figure}

%Here we use the power-law formula {[}See Eqs.~(\ref{Eq:imaginaryselfenergy}) and (\ref{Eq:sigma1}) in the main text{]} to fit the numerical results for Gaussian correlated disorder potential:
%\begin{equation}
%	\begin{aligned}\Sigma(E)=D\mathrm{sgn}(E)|E|^{1-\beta}+CE-i(\Sigma_{0}+\Delta|E|^{1-\beta}).\end{aligned}
%	\label{guassianfitting}
%\end{equation}
%We give the curve fitting results for different disorder strength for $\xi=2.0a$. The
%fitting parameters $\beta$ and $\Delta$ as a function of disorder strength ($\gamma$)
%are shown in Fig. \ref{fig:parameter}. As shown, linear fitting of $\beta$ and
%$\Delta$ with the disorder strength $\gamma$ give $\beta=2.04\gamma$ and $\Delta=0.42\gamma$.
%As shown in Fig. \ref{fig:parameter}, the dimensionless parameter $\Sigma_0\xi/v_f$ also has such feature and can be roughly superimposed into the same curve: $\Sigma_0=0.17v_f/\xi exp(-0.69/K)$.


\begin{thebibliography}{10}
\bibitem{note1}In general, the self-energy depends on the energy $E$ and the wavevector
	$\vec{k}$. However our simulation shows that the self-energy is only dependent on
	the energy.
\end{thebibliography}
\end{document}